%% file: qYMnetwork.tex
\pdfoutput=1
\documentclass[a4,12pt]{article}
\usepackage{comment}
\input{preamble.tex}
\begin{document}

\begin{titlepage}

\begin{flushright}
IPMU 16-0040
\end{flushright}
\vskip 2cm

\begin{center}

{\Large \bfseries
Wilson punctured network defects in 2D $q$-deformed Yang-Mills theory
}

\vskip 1.2cm

Noriaki Watanabe$^{\sharp}$

\bigskip
\bigskip

\begin{tabular}{ll}
$^\sharp$  & Kavli Institute for the Physics and Mathematics of the Universe, \\
& University of Tokyo,  Kashiwa, Chiba 277-8583, Japan
\end{tabular}

\vskip 1.5cm

\textbf{Abstract}
\end{center}

\medskip
\noindent
In the context of class S theories and 4D/2D duality relations there, we discuss the skein relations of general topological defects on the 2D side which are expected to be counterparts of composite surface-line operators in 4D class S theory. Such defects are geometrically interpreted as networks in a three dimensional space.
We also propose a conjectural computational procedure for such defects in two dimensional $SU(N)$ topological $q$-deformed Yang-Mills theory by interpreting it as a statistical mechanical system associated with ideal triangulations.

\bigskip
\vfill
\end{titlepage}
\setcounter{page}{1}

\tableofcontents

\pagebreak

%We compare the OPE between certain class of 4D surface defects  skein relations and

% any closed Wilson  network operators 

% a conjectural expression for any closed Wilson network operators in two dimensional $SU(N)$ topological $q$-deformed Yang-Mills theory.
%We propose a conjectural expression for any closed Wilson network operators in two dimensional $SU(N)$ topological $q$-deformed Yang-Mills theory.
%which is expected to be a geometrical description of surface-line system in class S theories.

%we can naturally extend the result to the finite area case but we do not know whether such a expression is correct or not...
%possible to extend but do not discuss here

\section{Introduction}

Recently, many interacting superconformal field theories (SCFTs) have been discovered whose definitions based on Lagrangians are not known yet.
In particular, there is a certain group of 4D theories often called ``class S theory" which are obtained as twisted compactifications of the 6d $\SUSYN{(2,0)}$ SCFTs on Riemann surfaces $C$ with punctures \cite{Gaiotto0904,GMN2,book:Tachikawa}.
Interestingly, even though almost all of the SCFTs have no definition based on the Lagrangians, some of their BPS observables have been evaluated assuming the dualities following their geometrical constructions.
In particular, for theories with $\SUSYN{2}$ Lagrangian descriptions, their partition functions on the squashed four-sphere $S^4_b$ \cite{Pestun07,HamaHosomichi1209}
and the superconformal indices (SCIs) \cite{KMMM05} (in the Schur limit) or equivalently partition functions on $S^1 \times_q S^3$ \cite{Romelsberger05} were computed.
Based on their explicit expressions, it was recently suggested that many class S theories beyond the Lagrangian definition
\footnote{In this paper, we focus on the Riemann surface compactifications with more than two regular punctures and no irregular ones. There are generalized proposals for non-SCFTs and Argyres-Douglas theories which need such irregular punctures \cite{GMN2,Gaiotto0908,GaiottoTechner12,BuicanNishinaka1505,CordovaShao1506,Song1509}.}
have alternative effective descriptions by some 2D theories : Liouville/Toda CFTs for $S^4_b$ case \cite{AGT09,Wyllard09} and 2D topological $q$-deformed Yang-Mills for $S^1 \times_q S^3$ case \cite{GRRY1104,GRRY1110,GPRR0910}.
%In particular, the conjectural SCI expression of the Argyres-Seiberg dual theory \cite{ArgyresSeiberg07} for $SU(3)$ $N_f=6$ SCFT was obtained using the inversion formula \cite{}.
%\footnote{because three point function of Toda conformal field theories are undetermined \cite{Fateev-L...} But there is a totally different approach \cite{}}
%They proposed that
Indeed, in addition to the partition functions, these 4D/2D dualities also offer new geometrical descriptions of supersymmetric defects in such SCFTs, which are main subjects of this paper.
%free from the perturbative definitions 

First, let us focus on the 4D gauge theory.
In particular, there are supersymmetric Wilson-'t Hooft line operators \cite{Wilson74,'tHooft78,Maldacena9803,Kapustin05,KapustinWitten06,GOP11,DGG1112,GKL1201,IOT11}
and half-BPS surface operators \cite{GukovWitten06,GukovWitten08,Nakayama1105,Gaiotto0911,DGLL12,GaddeGukov13,Gukov1412}.
It is considered that both defects come from codimension-four defects appearing in 6D $\SUSYN{(2,0)}$ SCFTs
\footnote{Precisely speaking, some surface operators can also come from codimension-two defects in the 6D theory \cite{AldayTachikawa1005,KPPW10,Tachikawa1102,FGT15}. However, we do not pay attentions to them in this paper.}
and both have the same origin in 6D. But their appearances in those 2D theories on $C$ look totally different.
The 4D line operators correspond to Verlinde network operators/Wilson network operators in the Liouville-Toda CFTs/$q$-deformed Yang-Mills theories, see \cite{DMO09,Xie1304,Xie1312,CGT15} for the geometrical viewpoint, \cite{Verlinde88,AGGTV09,DGOT09,Passerini10,GomisLeFloch10,DGG10,Bullimore13} for the Verlinde network and \cite{Migdal75,Witten89NPB322,Witten89NPB330,CMR94,BuffenoirRoche94,AOSV04,TW1504} for the Wilson network.
However, there is a serious problem left : How can we compute the expectation values of general network defects in the 2D $q$-deformed Yang-Mills theories ?
\footnote{See \cite{Bullimore13} for the Verlinde networks.}
Naively speaking, it seems to be enough to replace ordinary Lie groups by ``quantum group" as gauge groups at mathematical level. Indeed, the rigorous definition of Wilson loops without junctions in that case was given in \cite{BuffenoirRoche94} based on quantum groups, but its extension to any networks is not obvious yet for several reasons. Furthermore, even if it can be well-defined, it is not useful for the actual computations because it needs the general invariant tensors in the quantum group sense.
Instead of giving rigorous definitions, we will propose the direct procedure to obtain the conjectural expressions in Sec.~\ref{sec:closed Wilson network}. %This is the half of the aims in this paper.
The most important evidence for this proposal is the reproduction of several skein relations as remarked later.

On the other hand, the 4D surface operators are mapped into (fully degenerate) vertex operators/difference operators in the CFTs/Yang-Mills theories, see \cite{AGGTV09,DGG10,GomisLeFloch14} and \cite{GRR1207,GaddeGukov13,ABFH13,BFHR14,ABF13,RazamatYamazaki13}.
Geometrically, they can be represented as special punctures on the Riemann surface in both the set-ups.
Here let us consider a situation in which both line operators and surface operators coexist \cite{GMN4}.
This is the main subject in this paper.

There arise two questions : 1. Can we describe their skein relations on the geometrical side ? \. 2. How can we extend the previous conjectural formula for closed Wilson network defects to these more general cases ?
A key observation to answer these questions is as follows.
On both the Liouville/Toda CFT and the $q$-deformed Yang-Mills theory, the concept of crossings of networks exists and, in fact, they correspond to the ordering of corresponding half-BPS line operators in one space direction determined by the unbroken supersymmetry in 4D gauge theory \cite{GMN3,IOT11,GKL1201,TW1504}.
Then, the existence of crossings among several networks suggests that there is a hidden direction which is exactly identified with one of physical directions on the 4D gauge theory side.
\footnote{If we replace the 4-manifold on which the gauge theory is defined by the squashed 4-sphere $S^4_b$, there are locally two such directions which are exchanged under the flip from $b$ to $b^{-1}$ \cite{HamaHosomichi1209}. Here we focus on either direction.}
In other words, there appears a three dimensional geometry combined with the 2D space $C$ and one of 4D directions which is determined by the unbroken supersymmetry for the half-BPS loop operators.
This is the more familiar story in RCFTs whose conformal blocks are the wave functions of the corresponding 3D Chern-Simon theories \cite{Witten88-Jones,Witten89-complex}.
See also \cite{WuYang09,Gaiotto1404} as for Verlinde loop operators in the Liouville CFT. We must note that the closer story exists for $q$-deformed Yang-Mills theory \cite{AOSV04,deHaro0412,SzaboTierz1305} but we do not know the precise relation between two systems.
When we recall that the expectation values of BPS loops are independent of the positions on that direction \cite{GMN3,HamaHosomichi1209,IOT11,GKL1201}, it is natural to speculate that the networks are still topological in the new geometry.
In this new three dimensional geometry, codimension four defects are expressed as knot with junctions and both surface defects and line defects are on the same ground. We refer the corresponding defects in the $q$-deformed Yang-Mills theory to as ``punctured networks", which are the central subjects in Sec.~\ref{sec:composite} and describe the answer to the first question.

%On the other side, it is possible to add the 4D surface operators originating from codimension two defects in 6D $\SUSYN{(2,0)}$ SCFT.
%Since they are extended along that direction and punctures on the compactified Riemann surface, their locus can be considered as one-dimensional object in the three dimensional geometry.
%At this stage, we have one dimensional objects for both surface defects and line defects.
%This perspective is natural in term of 6D $\SUSYN{(2,0)}$ SCFT because both defects come from the codimension two defects.

In Sec.~\ref{sec:punctured network}, we combine the two results in previous two sections and give the expectation values formula for general punctured networks, which is the answer to the above second question.
In the Appendix.~\ref{app:more math}, we develop the method to evaluate the proposed formula in Sec.~\ref{sec:closed Wilson network} and summarize some mathematics appearing there. Using them, we prove that our proposal indeed reproduces a few fundamental skein relations in the following Appendix.~\ref{app:derivation of skeins}. This check gives a strong evidence that our proposal of the formula works well.
The Appendix.~\ref{app:charge_network} is the complement of charge/network correspondence discussed in \cite{TW1504}. We can expect the same structure for the composite surface-line systems.

\input{closed_network.tex}

\input{composite_defects.tex}

\input{open_network.tex}

\section{Summary and discussion}

In this paper, we propose the conjectural computational procedures for the general closed and punctured network defects in the 2D $q$-deformed Yang-Mills theory.
Such networks are geometrically knots with junctions in the three dimensional space which includes $C$ and one of 4D directions and expected to be the counterparts of composite surface-line systems in 4D.

Now we list several problems to be solved in future.
In the gauge theory perspective, it is necessary to discuss the 4D descriptions of the composite surface-line systems and compare the SCIs with the expectation values.
The 4D line operators are bounded to the surface operators and expected to be some interfaces including 2D Wilson lines of the two dimensional $\SUSYN{(2,2)}$ gauged linear sigma models \cite{HoriRomo1308,HondaOkuda1308,SugishitaTerashima1308}.
On the other hand, since the three geometry is encoded in 5D space, it is possible to describe them based on 5D SYM language like \cite{FKM12}.
It is also interesting to relate them to the well-known 3D-3D correspondence story \cite{DGG1108,DGG1112} where the 3D $\SUSYN{2}$ gauge theories on $S^3$ and complex Chern-Simons theories on hyperbolic spaces are related. More additions of defects in this correspondence were also discussed in \cite{GKRY15}, for example.

There are still several generalizations in this set-up.
One thing is to define and to incorporate general open networks \cite{Gaiotto1404} which are composite systems of codimension two and four defects.
Other is the extension to general simple Lie algebras (simply-laced in the context of class S), in particular $D$-series \cite{LPR12,CDT1601} or in the presence of twisted lines in $A_{2N-1}$-series \cite{CDT1212}.
Finally, although it is expected that the finite area extension is straightforward \cite{Tachikawa1207}, are there interesting applications ?

There are also several mathematical problems :
the justification of global symmetry enhancement
in all order $q$ expansion in Sec.~\ref{subsec:examples},
skein relations in the presence of general punctures,
reproduction of $q=1$ limit results where there is the unambiguous definition of Wilson networks, more on quantum groups behind \cite{CGR1507} and relations to integrable models as remarked in Sec.~\ref{subsec:remark on R-matrix}.
The relation to higher Teichm\"{u}ller space structure  \cite{Xie1304} or Liouville-Toda analysis \cite{Teschner1412,VartanovTeschner13} is also interesting because the local information of OPEs are expected to be same in both $S^4_b$ and $S^1\times_q S^3$ systems.

\paragraph{Acknowledgements}
The author would like to gratefully thank Yuji Tachikawa for many discussions on trying to define Wilson networks based on quantum group and making several comments on his draft.
He also wishes to thank the kind hospitality by the staffs of Perimeter Institute for theoretical physics where he had finished writing this draft, and great members there, in particular, Jaume Gomis, Davide Gaiotto, Shota Komatsu and Hee-Cheol Kim, for making his visit fruitful and several discussions.
He would like to thank Takuya Okuda, Masahito Yamazaki and Kentaro Hori for various discussions on this theme and finally sofas at Kavli IPMU and Perimeter Institute for relaxing him to advance this research.
The author is supported by the Advanced Leading Graduate Course for Photon Science, one of the Program for Leading Graduate Schools lead by Japan Society for the Promotion of Science, MEXT and also the World Premier International Research Center Initiative (WPI), Kavli IPMU, the University of Tokyo.

\appendix

\input{appendix.tex}

\bibliographystyle{JHEP}%alpha
\bibliography{add,qYM,defect,defectAGT,books,classS,AGT,math}

\end{document}

%% file: preamble.tex
\usepackage{amsmath,amsthm,amssymb}
\usepackage{graphicx}
\usepackage[usenames,dvipsnames]{xcolor}
\definecolor{MyGreen}{rgb}{0,0.5,0.2}
\colorlet{green}{MyGreen}
\usepackage{wrapfig}
\usepackage{enumerate}
\usepackage{mathrsfs}
\usepackage{geometry}
\usepackage{fancyhdr}
\usepackage{caption}
\usepackage{fancybox}
\usepackage{youngtab}

\usepackage{hyperref}
\usepackage{cite}
\usepackage{bm}
\usepackage{braket}

\allowdisplaybreaks

\usepackage{tikz}
\input{tikz.tex}

\numberwithin{equation}{section}
\def\SUSYN#1{{\mathcal{N}{=}#1}}
\def\GNOL#1{#1^\vee}

\def\wtl#1{\widetilde{#1}}
\def\ovl#1{\overline{#1}}

\def\sgn{{\rm sgn}}

\def\bbZ{\mathbb{Z}}

\def\bbR{\mathbb{R}}
\def\bbC{\mathbb{C}}

\def\bbT{\mathbb{T}}

\def\Lieh{\mathfrak{h}}

\def\Liesu{\mathfrak{su}}

\def\frq{\mathfrak{q}}
\def\calE{\mathcal{E}}
\def\calF{\mathcal{F}}

\def\calG{\mathcal{G}}
\def\scrG{\mathscr{G}}

\def\calO{\mathcal{O}}
\def\scrO{\mathscr{O}}

\def\calR{\mathcal{R}}

\def\scrS{\mathscr{S}}

\def\calW{\mathcal{W}}

\def\calZ{\mathcal{Z}}

\def\for{\quad \mbox{for} \;\;}

\def\subul#1#2{#1_{\underline{#2}}}

%% file: tikz.tex
\usetikzlibrary{calc}
\usetikzlibrary{arrows,shapes,positioning}
\usetikzlibrary{decorations.markings}
\usetikzlibrary{decorations.pathreplacing}
\usetikzlibrary{mindmap,trees,shadows}
\usetikzlibrary{patterns}
\usetikzlibrary{intersections}

\usetikzlibrary{external}
\tikzstyle no arrow=[thick]
\tikzstyle end arrow=[thick,postaction={decorate,decoration={markings, mark=at position .999 with {\arrow[scale=1.4]{stealth}}}}]
\tikzstyle mid arrow=[thick,postaction={decorate,decoration={markings, mark=at position .6 with {\arrow[scale=1.4]{stealth}}}}]
\tikzstyle mid-end arrow=[thick,postaction={decorate,decoration={markings, mark=at position .8 with {\arrow[scale=1.4]{stealth}}}}]
\tikzstyle start-mid arrow=[thick,postaction={decorate,decoration={markings, mark=at position .35 with {\arrow[scale=1.4]{stealth}}}}]
\tikzstyle latter arrow=[thick,postaction={decorate,decoration={markings, mark=at position .8 with {\arrow[scale=1.4]{stealth}}}}]
\tikzstyle former arrow=[thick,postaction={decorate,decoration={markings, mark=at position .4 with {\arrow[scale=1.4]{stealth}}}}]
\tikzstyle transarrow=[white,thick,postaction={decorate,decoration={markings, mark=at position .0 with {\arrow[scale=1.4,black]{stealth}}}}]

\newcommand{\circlearrow}[6]{\draw[transarrow] ($#1+(#3:#2)+(#3+90:#4*#2)$) node[black,#5]{#6} --++ (#3+90:#2);}
\newcommand{\anticirclearrow}[6]{\draw[transarrow] ($#1+(#3:#2)+(#3-90:#4*#2)$) node[black,#5]{#6} --++ (#3-90:#2);}

%% file: closed_network.tex
\section{Closed Wilson network defects in 2D $q$-deformed Yang-Mills}
\label{sec:closed Wilson network}

In this section, we see how the expectation values of any closed Wilson network defects can be evaluated.
Here ``closed" means that the networks do not touch on general punctures coming from codimension two defects in 6D SCFTs not codimension four ones.
We divide the evaluation procedure into two steps : giving the computational procedures for special cases (See Sec.~\ref{subsec:proposal}) and then constructing the general cases by using them (See Sec.~\ref{subsec:reduction}).
However, we will explain these two steps in the reversed order by starting the general cases and then by decomposing them into several special building blocks for which we will give the procedure.

The organization of this section is as follows.
In Sec.~\ref{subsec:reduction}, we show the procedure to obtain the special building blocks from the general set-ups.
Next, in Sec.~\ref{subsec:review on qYM}, we review some properties of 2D $q$-deformed Yang-Mills theory and class S theories needed later.
In Sec.~\ref{subsec:proposal}, we go back to the evaluation of defect expectation values. There, we map such evaluations for special building blocks into the computations of partition functions of statistical mechanical systems with infinite degrees of freedom. The construction of such a mapping and giving the Boltzmann factor are the main points.
We see some applications to a few concrete theories in Sec.~\ref{subsec:examples} and make a few comments on the above mapping of $\calR$-matrix in a special case in Sec.~\ref{subsec:remark on R-matrix}.

First of all, we recall some notations and properties needed in this paper. See also App.~\ref{app:more math} as for the Lie algebra notation.
We assume the following properties for closed networks (See \cite{Xie1304,Bullimore13} and \cite{TW1504} for the details.) :
\begin{enumerate}
\item Each network consists of trivalent junctions and arrowed edges with a charge $a \in \{0,1,2,\ldots,N-1\} \simeq \bbZ_N$ on each.
Each charge corresponds to the fundamental representations $\wedge^a \Box$ of $SU(N)$ which form the minimal set to generate all the irreducible representations.
Notice also that edges with $0$ can be removed and we ignore them hereafter.

Flipping arrow is equivalent to the replacement by the charge conjugate representation
\begin{align}
\raisebox{-0.3\height}{\input{figure/sec2/right_arrow.tikz}}
=\raisebox{-0.3\height}{\input{figure/sec2/left_arrow.tikz}}.
\label{eq:flip of arrow and cc of rep}
\end{align}
In particular, as we only consider the fundamental representations $\wedge^a \Box$, this operation corresponds to $a \to N-a$.

If we use an edge labelled by an irreducible representation $R$, we interpret it as a bunch of edges according to a polynomial expression of $R$ of $SU(N)$ representation ring generators of $\wedge^a \Box$.

\item There is the charge conservation on each junction. More precisely, if we have all three inflowing/outgoing edges with charges $a,b$ and $c$, they must satisfy $a+b+c = 0 \mod N$.
On forgetting to take the $N$-modulo operation, there are two possibilities : $a+b+c=N$ or $2N$.
\begin{figure}
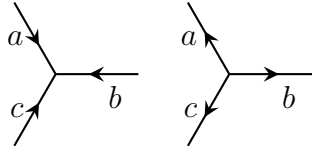

\centering
\input{figure/sec2/noncanonical1.tikz}
\input{figure/sec2/noncanonical2.tikz}\\
\caption{Inflowing (Left) and outflowing (right) $(a,b,c)$-junctions. \label{fig:non-canonical}}
\end{figure}
We call the former one $(a,b,c)$-junction for both inflowing one and outflowing one, see Fig.~\ref{fig:non-canonical}.
If $a+b+c=2N$, the redefinition $a'=N-a$, $b'=N-b$ and $c'=N-c$ makes $a'+b'+c'=N$ and the exchange between inflowing and outflowing, and we have $(N-a,N-b,N-c)$-junction for the latter case.

\item Any crossing can be resolved into a network without any crossing. In particular, in \cite{TW1504}, we identified such relations in class S theories, referred to as ``crossing resolutions", with those already found in \cite{MOY2000} like
\footnote{\label{fn:convention} There is an important caution. As explained in \cite{TW1504}, there are two different conventions called ``Liouville-Toda" convention and ``$q$-deformed Yang-Mills" one. Although we focus on the 2D $q$-deformed Yang-Mills theory, we also use the former convention which is mostly used in the context of the 4D/2D duality, after this section. There, instead of $q$, we use another symbol $\frq$ which is related to $q$ by $\frq=q^{1/2}$. Note also that the skein relations in the $q$-deformed Yang-Mills convention are obtained under the replacement of $\frq$ by $-q^{1/2}$, where the additional minus sign appears compared to the above actual relation. This is because the normalizations of the junctions differ in two conventions.}
\begin{eqnarray}
\raisebox{-0.4\height}{\input{figure/sec2/general_R.tikz}}
=
\frq^{\frac{ab}{N}}
\sum_{i=0}^{s}
\frq^{-i}
\raisebox{-0.5\height}{\input{figure/sec2/general_Q.tikz}}
\label{eq:conjectural skein relation of general crossing}
\label{eq:final form of conjectural skein relation of general crossing}
\end{eqnarray}
where $s=\min(a,b,N-a,N-b)$ and $a,b=0,1,2,\ldots,N-1$.
On the right hand side, there are no crossings. Therefore, we can remove all the crossings from the network by applying the above relation to each but have a sum of several networks instead.
\end{enumerate}

\subsubsection*{Skein relations}

Once we have introduced the 3D geometry discussed in Sec.~\ref{subsec:geometrical configurations}, the meaning of skein relations are exactly same as those in the knot theory.
Instead, throughout this paper, we use the skein relations in the following sense.

Let $W_q[\Gamma](\{ a \})$ be the expectation value of the 2D Wilson network operator associated with $\Gamma$ in the 2D $q$-deformed Yang-Mills theory. $\{ a \}$ are all holonomies around punctures of $C$.
And let us consider two sub graphs $\gamma_A$ and $\gamma_B$. For any pair of two graphs $\Gamma_A$ and $\Gamma_B$ which include $\gamma_A$ and $\gamma_B$ respectively but are same on removing these sub graphs,
when the equality shown just below always holds true, we identify $\gamma_A$ and $\gamma_B$ and write this as $\gamma_A \sim \gamma_B$. The equality is
\begin{align}
W_q[\Gamma_A](\{ a \})=C_q(\gamma_A \to \gamma_B)W_q[\Gamma_B](\{ a \})
\end{align}
where $C_q(\gamma_A \to \gamma_B)$ is a function of only $q$ and independent of all holonomies around punctures and is also determined by $\gamma_A$ and $\gamma_B$ only.
\footnote{In all examples we know, $C_q$ is a product of a polynomal of $q^{\frac{1}{2}}$ and a monomial with a negative rational power of $q$.} Notice that, in many cases, this relation is enough local and independent of the choice of $C$.
Now we have the equivalence relations $\sim$ and refer to them as the "skein relations" hereafter.

Finally, we make a few comments.
Under the parameter identification $q=e^{2\pi i b^2}$ \cite{DGG1108,TW1504} where $b$ is a physical parameter in the Liouville/Toda CFT, the skein relations are common both in the CFTs and in the 2D $q$-deformed Yang-Mills theory.
This is because the skein relations are expected to be the local relations about codimension four defects in the 6D $\SUSYN{(2,0)}$ SCFTs and to be independent of the four dimensional global background geometries, namely, the choice of $S^4_b$ or $S^1 \times_q S^3$.

Notice also that the crossing skein relation may suggest the new direction of the 2D $q$-deformed Yang-Mills theory but the appearance is not so obvious in this 2D theory itself.
However, this class S picture from the 6D $\SUSYN{(2,0)}$ SCFTs strongly suggest that. See also Sec~\ref{subsec:geometrical configurations}.

\subsection{Reduction onto special cases}
\label{subsec:reduction}

Here we see how the most general pairs of the Riemann surface and defects on it decompose into the several special ones as the building blocks.

\begin{comment}
special class for pairs of a Riemann surface and Wilson network defects on it.

Let $C$ be a Riemann surface with genus $g$ and $n$ punctures and $\Gamma$ be all closed networks on it.
Any network consists of three kinds of components : loop around punctures, loop wrap a tube (including punctures) and networks with several trivalent junctions
\footnote{Mathematically speaking, the homology class of the former loops belong to the kernel of the natural inclusion map from $H(C,\bbZ)$ to $H(\bar{C},\bbZ)$ obtained by removing the punctures. In physics language, they are just flavor Wilson loops coupled to the background gauge fields.}

The reduction consists of the several steps as follows.

\begin{enumerate}
\item removal of loops

After applying the crossing resolutions, we can decompose $C$ into several components $C_{\alpha}$ by cutting along the purely loops without junctions.
Then we have the pair $C_\alpha$ and the closed networks $\Gamma_\alpha$ on each if we regard the cut sections as new punctures.

\item

\end{enumerate}

\end{comment}

\begin{figure}[t]
\centering
\includegraphics[width=100mm]{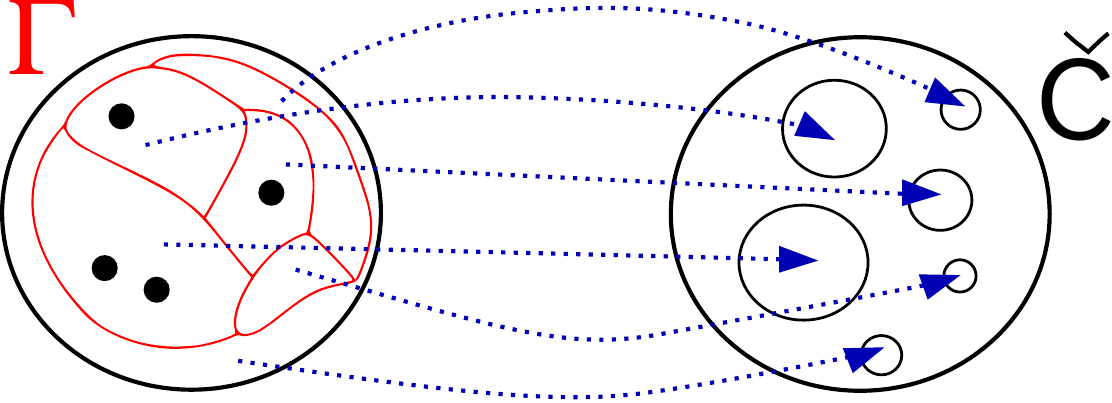}
\caption{A fat graph $\check{C}$ from a network $\check{\Gamma}$. The red graph represents the network and each region is mapped to a hole. In this example, this is isomorphic to the six-punctured sphere.}
\label{fig:fat graph}
\end{figure}

Let $C$ be a Riemann surface with genus $g$ and $n$ punctures and $\Gamma$ be any closed networks on it.
To each puncture, we assign a holonomy which corresponds to the fugacity in the SCI language. On the types of punctures and their holonomies, see the review in Sec.~\ref{subsec:review on qYM} later.
$\Gamma$ may consist of several disconnected components and we write the decomposition as $\Gamma=\underset{\alpha}{\sqcup}\check{\Gamma}_\alpha$.
Next, consider a neighborhood of $\check{\Gamma}_\alpha$ which is sometimes called a ribbon graph or a fat graph.
This fat graph denoted by $\check{C}_\alpha$ is a two dimensional open surface and its boundary consists of several copies of $S^1$. See Fig.~\ref{fig:fat graph}.
Cutting along the boundaries of $\check{C}_\alpha$, we have a decomposition of $C$.
By the above construction, in addition to $\check{C}_\alpha$'s, there are other connected components denoting $\wtl{C}_A$ which have no network defects.
Note that we identify each boundary isomorphic to $S^1$ with a puncture.

Let us make one comment on the topological property of $\check{C}$.
If $\check{\Gamma}$ has $2\ell \;(\ell > 0)$ junctions and the boundary of its fat graph $\check{C}$ is isomorphic to $k$ copies of $S^1$, $\check{C}$ is the $k$-punctured genus $(\ell-k)/2+1$ Riemann surface. Note that its Euler character is $-\ell$.
\footnote{\label{fn:Euler number for special case} Let $e$,$v(=2\ell)$ and $f$ be the number of edges, junctions of $\check{\Gamma}$ and regions in $\check{C}$, respectively. By construction, $f=k$ holds true.
The closedness of the network $\check{\Gamma}$ and the trivalence property of junctions also say $2e=3v$.
The Euler's theorem applied to $\check{C}$ ignoring all the punctures gives $2-2g=f-e+v$.
Combined with all, we finally have the claim $\chi_{\check{C}}=2-2g-k=-\tfrac{1}{2}v=-\ell$.}
In particular, when $\check{\Gamma}$ is a pure loop without junctions, $\check{C}$ is the twice-punctured sphere.

Now $C$ consists of two types of connected components : $\check{C}_\alpha$ which is homotopic to $\check{\Gamma}_\alpha$ and $\wtl{C}$ for which we already know how to compute their partition functions as remarked later.
Since we can reconstruct the expectation values of the original system by gluing together as shown around ~\eqref{eq:gluing}, all we have to do is know the expectation values for each pair $(\check{C}_\alpha,\check{\Gamma}_\alpha)$.
Before showing that procedure (Sec.~\ref{subsec:proposal}), we review several facts needed for the complete reconstruction and later discussions.

\subsection{Brief reviews on the partition functions and loops expectation values}
\label{subsec:review on qYM}

In this section, we review the following five points :
\begin{enumerate}
\setcounter{enumi}{-1}
\item Overall renormalization factors
\item Formula for no network defect cases
\item Gluing (= gauging process on the 4D side)
\item Formula in the presence of loops
\item Partially Higgsing or partially closing on punctures.
\end{enumerate}
See \cite{AOSV04,deHaro0509} on the $q$-deformed Yang-Mills, and see also a review \cite{CMR94} when $q=1$. Refer to \cite{Gaiotto0904,GRRY1104,Tachikawa1504} on the class S punctures.

Here we define $q$-number as
\begin{align}
[n]_q:=\dfrac{q^{n/2}-q^{-n/2}}{q^{1/2}-q^{-1/2}}.
\end{align}
$\dim_q R$ denotes the $q$-dimension of an irreducible representation $R$. See ~\eqref{eq:q-dimension} for its definition.
If we want to change the convention from $q$-deformed Yang-Mills one to Liouville-Toda one, all we have to do is to replace the above $q$-number by quasi $\frq$-number
\begin{align}
\langle n\rangle_\frq:=(-1)^{n-1}\dfrac{\frq^{n}-\frq^{-n}}{\frq-\frq^{-1}}.
\end{align}

\subsubsection*{0. Overall renormalization factors}

On the comparison between the SCIs in the Schur limit and $q$-deformed Yang-Mills partition functions, there is a bit difference between two :
There are additional overall factors in the SCIs. It consists of two types : a prefactor depending only on $q$ and renormalization factors $K(a,q)$ assigned to punctures.
The former factor is irrelevant in our discussion and we neglect it.
The latter one is given by the inverse square root of the SCI of $SU(N)$ free vector multiples.
The $q$-deformed Yang-Mills partition functions are obtained by removing this factor for each puncture from the SCI expressions and by replacing each vector contribution by $1$ which can be interpreted as the integral measure over the fugacity.
Since $\displaystyle \lim_{q\to 0} K(a,q)=1$, this factor can also be ignored as long as we see the leading order in the $q$-expansion of the expressions, which is exactly done in Sec.\ref{subsec:examples} in this paper. We need the concrete expressions of such factors if we compute them at the higher order in the $q$-expansion.

\subsubsection*{1. Formula for no network defect cases}

When $\wtl{C}$ is a genus $\wtl{g}$ Riemann surface with $\wtl{n}$ punctures with $SU(N)$ holonomies,
\footnote{In the language of class S theory, they are called maximal (or full) punctures.}
 its partition function $I_{\wtl{C}}(\{z\})$ is given by
\begin{eqnarray}
I_{\wtl{C}}(\{z\})=\sum_{R} (\dim_q R)^{\chi_{\wtl{C}}} \prod_{i=1}^n \chi_{R}(z_i)
\label{eq:qYM p.f. without defects}
\end{eqnarray}
where $R$ runs over all unitary irreducible representations of $SU(N)$, $\{z\}$ represents the set of holonomies and $i$ is the index of punctures.

For each summand labelled by $R$, we can represent the Riemann surface with which $R$ is assigned.
This interpretation will be important later.

\subsubsection*{2. Gluing}

There is a natural operation, gluing, which identifies two holonomies on different punctures and connect them geometrically. On the 4D SCFT side, there are two $SU(N)$ flavor symmetries which are identified by adding the vector multiplet \cite{GMT11}.
%More physically speaking in the context of 6D SCFTs, coupling $SU(N) \times SU(N)$ non-linear $\sigam$ model to them and 

\begin{figure}[t]
\centering
\includegraphics[width=100mm]{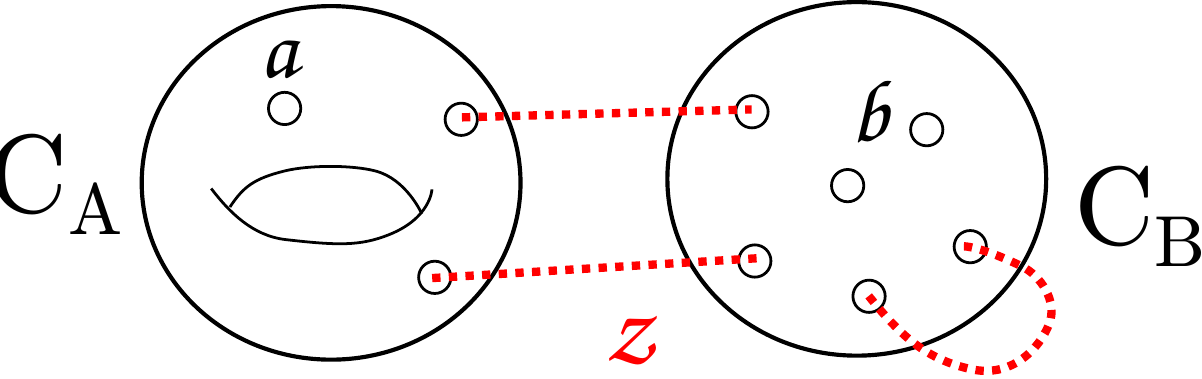}
\caption{Any Riemann surface can be constructed from more fundamental ones by gluing pairs of punctures as dashed lines.}
\label{fig:gluing}
\end{figure}

If we have two pairs of a Riemann surface with punctures and generic networks on it allowing the case of empty, which are denoted by $(C_A,\Gamma_A)$ and $(C_B,\Gamma_B)$, we can construct new one $(C_{AB},\Gamma_A \sqcup \Gamma_B)$ by gluing each pair of several punctures. See Fig.~\ref{fig:gluing}.
Let $I_{C,\Gamma}(\{z\})$ be the expectation values of the 2D topological $q$-deformed Yang-Mills theory on $C$ with a Wilson network defect $\Gamma$.
The corresponding expectation values can be constructed as
\begin{align}
I_{C_{AB},\Gamma_A \sqcup \Gamma_B}({\{a\},\{b\}})=\prod_i \left( \oint [dz_i] \right)
I_{C_A,\Gamma_A}(\{z^{-1}\},\{a\})
I_{C_B,\Gamma_B}(\{z\},\{b\})
\label{eq:gluing}
\end{align}
where $[dz]$ is the Haar measure of $SU(N)$, $\{z\}$ are gauged fugacities and $\{a\}$ and $\{b\}$ are ungauged ones on $C_A$ and $C_B$, respectively.
The independence of the order in the glues is obvious.

\subsubsection*{3. Formula in the presence of loops}

According to the result in \cite{DGG1112,GKL1201}, the SCI in the presence of 4D loop operators turns out to coincide with the VEV of Wilson loops in the 2D $q$-deformed Yang-Mills as discussed in \cite{TW1504}.
This can be obtained by simply adding the corresponding $SU(N)$ character on gluing as seen soon later.

\begin{figure}[th]
\centering
\includegraphics[width=100mm]{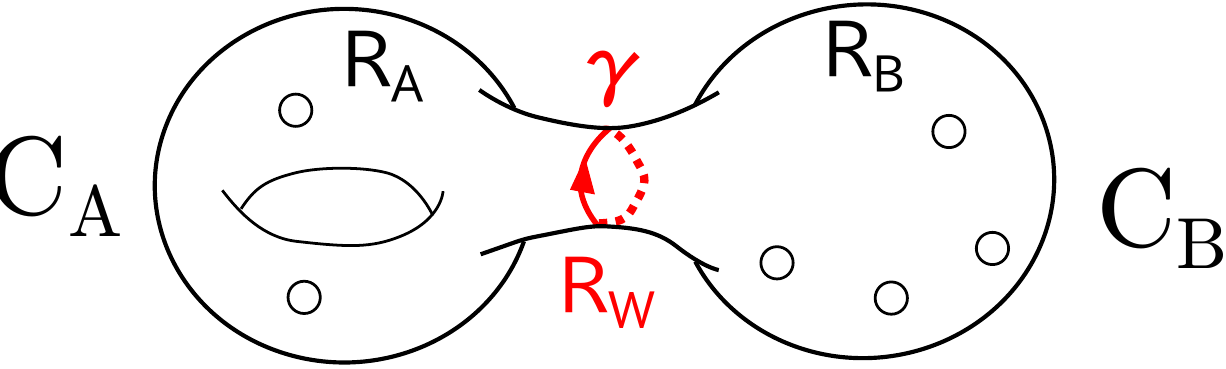}\\
\caption{A 2D Wilson loop in $R_W$ around the cylindrical part in $C$.}
\label{fig:loop in C}
\end{figure}

In this case, $\Gamma$ is a pure loop $\gamma$ along a one cycle in $C$ as depicted in Fig.~\ref{fig:loop in C}.
Let us cut along the Wilson loop labelled by an irreducible representation $R_W$, which is exactly the reversed operation to the previous gluing process, and assume that they are separated after the cut for simplicity.
\footnote{If not, it is enough to replace two expectation values $I_{C_A}$ and $I_{C_B}$ by a single one in ~\eqref{eq:loop expression in q-YM def}.}
Let $z$ denote the new holonomy or fugacity along the new boundary cycle.
Using the new Riemann surfaces $C_{A}$ and $C_{B}$ which have two additional punctures in total compared to $C$, we can express the Wilson loop expectation value of the 2D $q$-deformed Yang-Mills as
\footnote{The convention about the orientation adopted in this paper differs from \cite{TW1504}. This change is necessary when the surface operators are included because the orientation of the new direction is fixed. See the detail in later Sec.\ref{sec:composite}.}
\begin{align}
I_{C,\gamma}(\{a\})&=\oint [dz] \chi_{R_W}(z)
I_{C_{A}}(z^{-1},\{a\})I_{C_{B}}(z,\{b\})
\label{eq:loop expression in q-YM def} \\
&=\sum_{R_A,R_B} N_{R_B R_W}^{\quad\quad R_A}
(\dim_q R_A)^{\chi_{C_A}} (\dim_q R_B)^{\chi_{C_B}} \prod_{i} \chi_{R_A}(a_i)
\prod_{i} \chi_{R_B}(b_i)
\label{eq:loop expression in q-YM}
\end{align}
where $N_{R_B R_W}^{\qquad R_A}$ is the Littlewood-Richardson coefficient which counts the multiplicity of the representation $R_A$ appearing in the tensor product of $R_B$ and $R_W$.
The recursive applications can give the computations in all the general cases that $\Gamma$ consists of multiple loops and networks.

\begin{figure}
\centering
\input{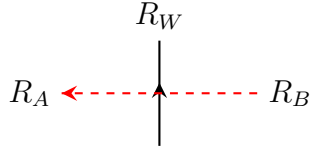}\\
\caption{Two adjacent regions across an edge labelled by $R_W$.}
\label{fig:neighbor_region_via_edge}
\end{figure}

If we have two isolated regions across an edge labelled by $R_W$, to each summand in ~\eqref{eq:loop expression in q-YM}, we can assign irreducible representations $R_A$ and $R_B$ to $C_A$ and $C_B$, respectively. See Fig.\ref{fig:neighbor_region_via_edge}.
The summand vanishes unless $N_{R_B R_W}^{\quad\quad R_A} \neq 0$ and then we have the constraint $R_B \in R_A \otimes R_W$ meaning that the irreducible decomposition of $R_A \otimes R_B$ includes $R_W$.
In particular, in our convention, the charge on each edge is a fundamental representation $\wedge^a \Box$ and the above constraint on $R_A$ and $R_B$ becomes powerful, which will turn out to be useful in the analysis in Sec.~\ref{subsubsec:allowed}.

Let us make a few remarks.
When $\calZ_R$ denotes the center charge of $R$, this constraint leads to $\calZ_{R_B} + a = \calZ_{R_A} \mod N$ when $R_W=\wedge^a \Box$.
This implies that the expectation values vanish unless all the intersection numbers of any one cycles
\footnote{We define it by summing up all the intersecting edge's charges flowing from the left to the right along the one cycle following its orientation.}
 with the Wilson networks vanish.
For example, in the case that $C$ is an once-punctured torus, the expectation value of the fundamental Wilson loop along $\alpha$-cycle vanishes.
In particular, when the puncture is special called simple or minimum, this introduces a Wilson loop in $\Box$ in some duality frames on the 4D $SU(N)$ gauge theory side but it is localized at a point in $S^3$ which is a compact space.
Its center charge is not screened by the dynamical matter because all belongs to the adjoint representations and this theory is anomalous because there is a single source with a non-trivial Abelian charge on the compact space \cite{book:fields-strings,GKSW14}.

\subsubsection*{4. Partially Higgsing/closing}

If we have global $SU(N)$ symmetry in $4D$ $\SUSYN{2}$ SCFTs, there are BPS primary operators in the same supermultiplet as the flavor current belongs to. They are the triplet of $SU(2)_R$ R-symmetry and the adjoint representation of $SU(N)$ global symmetry.
By giving a nilpotent VEV to the highest weight of those operators at UV point, we have another SCFTs in the IR. This is called partially Higgsing/closure operation \cite{CDT12,Tachikawa1504}. The nilpotent orbit of $SL(N,\bbC)$ can be always uniquely mapped to the Jordan normal forms $J_Y=\oplus_i J_{n_i}$ whose all eigenvalues vanish and they are classified by the partition $Y=[n_1,n_2,\ldots,n_d]$ of $N$.
Now that $SU(2)_R \times SU(N)$ global symmetry is spontaneously broken into a subgroup $U(1) \times G_{Y}$.
In particular, this $U(1)$ is generated by 
\begin{align}
I^3_{IR}:=I^3_{UV} \otimes 1 - 1 \otimes \dfrac{1}{2}\rho^3_Y
\end{align}
where $\rho_Y$ is the unique embedding homomorphism from $\Liesu(2)$ into $\Liesu(N)$ satisfying $J_Y=\rho_Y\left[\left(\begin{array}{cc}
0 & 1 \\
0 & 0 \\
\end{array}\right)\right]$, $\rho_Y^3:=\rho_Y\left[\left(
\begin{array}{cc}
1 & 0 \\
0 & -1 \\
\end{array}
\right)\right]$ and $I^3_{UV}$ is a diagonal $R$-symmetry generator of $SU(2)_R$.
This $R$-symmetry generator in new IR SCFT is enhanced to $SU(2)_R^{IR}$ and the SCI can be also defined there.
The RG invariance implies
\begin{align}
q^{-I_{IR}^3} a_Y=q^{-I_{UV}^3} a
\end{align}
where $a_Y$ is the fugacities in the Cartan subgroup of $G_Y$.
In conclusion, the partially closing operation on each maximal puncture is equivalent to the following replacement \cite{GRRY1104,GRR1207,Tachikawa1504} :
\begin{align}
a \longrightarrow q^{\tfrac{\rho^3_Y}{2}} a_Y.
\end{align}
As for the notation used here, see the beginning of Sec.~\ref{sec:composite}.
Hereafter, we use the transpose of $Y$ to specify the type of punctures. For example, $[N]$ represents the full (maximal) puncture and $[2,1^{N-2}]$ does the simple (minimum) puncture.

\subsection{A proposal for closed Wilson networks}
\label{subsec:proposal}

At this stage, any expectation value of any network defect is a function of holonomies for $SU(N)$ global symmetries on each maximal puncture.
Recall that we can always take each $SU(N)$ holonomy in the maximal torus $\bbT^{N-1}$ which is a $N$-tuple of $U(1)$ holonomies $a_1,a_2,\ldots,a_{N}$ with the constraint $a_1a_2 \cdots a_N=1$ but there left the ambiguity of its permutations.
The invariance under the permutations (or conjugacy actions of $SU(N)$) implies that the expectation values can be expanded with the characters of $SU(N)$ again and written as
\begin{align}
I_{\check{C},\check{\Gamma}}(\{a\})=\sum_{\{R_p\}} B_{\check{\Gamma}:\{R_p\}} \prod_{p=1}^{n} \chi_{R_p}(a_i).
\label{eq:VEV of networks as sum over irrep's}
\end{align}
where $\{R_p\}$ means that each $R_p$ runs over the set of the unitary irreducible representations of $SU(N)$. We also use the same $n$ as before for the number of maximal punctures on $\check{C}$.
As we have seen in ~\eqref{eq:qYM p.f. without defects}, for any 2D $q$-deformed Yang-Mills partition functions without defects, the coefficient $B$ in the character expansion is diagonal in $\{R_i\}$ and each component is given by $(\dim_q R)^{\chi_{\check{C}}}$. In other words,
\begin{align}
B_{:\{R\}}=(\dim_q R)^{\chi_{\check{C}}} \prod_{p=1}^{n} \delta_{R_p,R}
\end{align}
where $\delta_{R_p,R}$ gives $1$ when $R_p=R$ and $0$ otherwise.

Now our goal is to give the procedure for computations of $B_{\check{\Gamma}:\{R\}}$.
For that purpose, let us interpret this as a Boltzmann factor of a statistical mechanics on a lattice system defined by following three steps :
\begin{enumerate}

\item Make a dual ideal triangulation

The Riemann surface $\check{C}$ decomposes into the several components by removing the locus of network defects $\check{\Gamma}$.
The previous construction via fat graphs ensures the natural one-to-one correspondence between the components of regions and the boundaries/punctures.
Now consider the dual quiver on $\check{C}$ associated with the network $\check{\Gamma}$.
This is obtained when each region or puncture and network's edge are mapped into a vertex and an arrow with the charge, respectively.
At this stage, the summation in ~\eqref{eq:VEV of networks as sum over irrep's} says that there lives a discrete but infinite physical degree of freedom labelled by the irreducible representations of $SU(N)$ or the dominant weights on each vertex.
Any junction is mapped into a triangle because of the trivalence property of the network.
Note that similar operations already appeared many times in various contexts, see \cite{GMN2,Xie1304} for example.
So we have (ideal) triangulations with a charge on each edge, of $\check{C}$.
Two or three vertices of a triangle are allowed to be common at this stage but, in Sec.~\ref{subsubsec:allowed}, we will see that we can ignore such triangulations.
%If we have a triangle whose vertices are not all distinct, the Boltzmann factor vanishes and we can ignore such quiver diagrams.
% %This can be easily understood 
%center charge of the irreducible rep.
Notice also that the number of triangles is given by $-2\chi_{\check{C}}$ on recalling footnote.~\ref{fn:Euler number for special case}.

\item Consider the allowed configurations of dominant weights

The whole configuration space is the set of all maps from each quiver vertex to an irreducible representation or a dominant weight.
However, for many configurations, $B_{\check{\Gamma},\{R\}}$ in $~\eqref{eq:VEV of networks as sum over irrep's}$ vanishes as we will discuss in ~\ref{subsubsec:allowed} and we can restrict the range of the summation to the non-vanishing configurations.
%It equals to assign vanishing Boltzmann 
%weights to the forbidden configurations in the next step.

\item Give the Boltzmann factor for each configuration

As with the ordinary statistical mechanics such as Ising models, we assume the Boltzmann factor of a given configuration is the product of local Boltzmann factors over all the triangles.
In other words, the local Boltzmann factor denoted $B^{\triangle}_{\lambda_A,\lambda_B,\lambda_C}$ is a function on triples of dominant weights living on three vertices of a single triangle $\triangle$ and the total Boltzmann factor is given by
\begin{align}
B_{\hat{\Gamma}:\{R\}} = \prod_{\triangle} B^{\triangle}_{\lambda_{\triangle,A},\lambda_{\triangle,B},\lambda_{\triangle,C}}
\end{align}
where $\triangle$ runs over all the triangles on the ideal triangulation of $\check{C}$.
The concrete formula of $B^{\triangle}_{\lambda_A,\lambda_B,\lambda_C}$ will be given as ~\eqref{eq:final exp. for local Boltzmann factor} in Sec.~\ref{subsubsec:formula for local Boltzmann factor}.

\end{enumerate}

\subsubsection{Selection rules on dominant weights}
\label{subsubsec:allowed}

Here we exhaust all the configurations whose Boltzmann factors are non-vanishing. The strategy is same as that used in \cite{TW1504}.

\begin{figure}[t]
\centering
\input{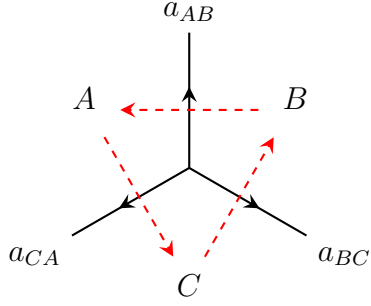}
\caption{There are three regions around each junction. The red dashed arrows represent its dual quiver.}
\label{fig:region and junction}
\end{figure}

Let us consider $(a_{AB},a_{BC},a_{CA})$-junction and three regions around it as shown in Fig.\ref{fig:region and junction}.
There are still two types of junctions, inflowing one and outgoing one but hereafter we focus on outflowing one only because the final expressions for $B^\triangle$ are same for both types.
Let three dominant weights living on its vertices be $\lambda_A,\lambda_B$ and $\lambda_C$ and define $\lambda_{XY}:=\lambda_{X}-\lambda_{Y}$ for $X,Y \in \{A,B,C\}$.
The gluing procedure stated in Sec.~\ref{subsec:review on qYM} tells that
$R(\lambda_Y) \otimes \wedge^{a_{XY}} \Box$ contains $R(\lambda_X)$ for $(X,Y)=(A,B),(B,C)$ and $(C,A)$.
This statement equals to $\lambda_{XY} \in \Pi(\wedge^{a_{XY}} \Box)$ where $\Pi(R)$ is a set of all weights of the highest representation $R$.
Then there is a unique subset $E_{XY}$ of $\{1,2,\ldots,N\}$ consisting of $a_{XY}$ elements such that $\displaystyle \lambda_{XY}=\sum_{s \in E_{XY}} h_s$.
The cycle condition $\lambda_{AB}+\lambda_{BC}+\lambda_{CA}=0$ means that $E_{AB}$,$E_{BC}$ and $E_{CA}$ has no common element and $E_{AB} \sqcup E_{BC} \sqcup E_{CA}=\{1,2,\ldots,N\}$.
In conclusion, allowed configurations have
several sectors determined by the choice of $E_{AB},E_{BC}$ and $E_{CA}$ which is a partition of $\{1,2,\ldots,N\}$ into three sets. Note that there are $\dfrac{N !}{a_{AB}! a_{BC}! a_{CA}!}$ sectors for single junction.
And in each sector, there is a summation over $\lambda_A$ for example,
\footnote{Of course, it is possible to choose $\lambda_B$ or $\lambda_C$ instead. In all cases, the other two dominant weights are determined if we specify the sector at first.}
 with a constraint that all $\lambda_A,\lambda_B$ and $\lambda_C$ are dominant weights.
\footnote{In other words, this is just summation over $\lambda_A$ and pairs of $\lambda_B-\lambda_A$ and $\lambda_C-\lambda_A$. The later two pairs label the sectors.}

Notice also that two adjacent vertices must have different center charges as we have seen at part 3 in Sec.~\ref{subsec:review on qYM} and we can say that there is no edge whose starting vertex and terminating one are common.

\subsubsection{Conjectural formula for the local Boltzmann factor}
\label{subsubsec:formula for local Boltzmann factor}

The last task is to show the way to get the local Boltzmann factor for any allowed triple of dominant weights $\lambda_A,\lambda_B$ and $\lambda_C$.
%There are at least two equivalent algorithms to compute the local Boltzmann factors but we give one way here and the other in App.~\ref{app:more math}.

\begin{figure}
\centering
\input{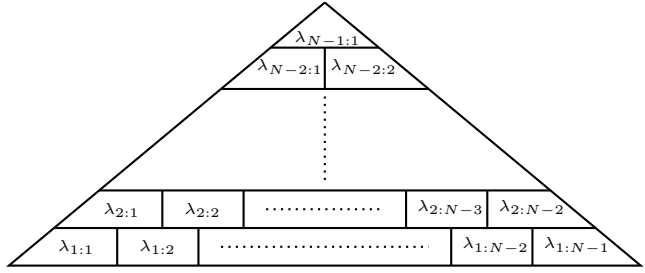}
\caption{Pyramid, an assembly of $\tfrac{1}{2}N(N-1)$ integers.}
\label{fig:pyramid}
\end{figure}

Before going to the final result, we must prepare some tools to express it simply.
First of all, we introduce a mathematical object playing central roles in our computations.
This is just an assembly of integers designated by two labels $h$ and $\alpha=\alpha_h$.
$h$ runs over $1,2,\ldots,N-1$ and $\alpha$ does over $1,2,\ldots,N-h$ for each $h$. Therefore, this object consists of $\tfrac{1}{2}N(N-1)$ integers.
We call such object ``pyramid" hereafter. See Fig.~\ref{fig:pyramid}.

In particular we have a natural map defined just below which sends a weight $\lambda=[\lambda_1,\lambda_2,\ldots,\lambda_{N-1}]$ where $\displaystyle \lambda=\sum_{\beta=1}^{N-1} \lambda_\beta \omega_\beta$ to a pyramid and denote the image by $\hat{\lambda}$ or $\hat{\lambda}_{h:\alpha}$.
The definition of the map is
\begin{align}
\hat{\lambda}_{h:\alpha} :=\sum_{\beta=\alpha}^{\alpha+h-1} \lambda_\beta.
\end{align}
Hereafter, we permit an abuse of notation. We use the same symbol $\hat{\lambda}$ for the pyramids not in the image of this inclusion map too. In such cases, $\hat{\lambda}$ is to be considered as a single symbol as a whole and $\lambda$ is meaningless.

Next, we define majority function $mj$ for three variables :
\begin{align}
mj(a,b,c):=
\begin{cases}
a \qquad b=a \;\textrm{or}\; c=a \\
b \qquad a=b \;\textrm{or}\; c=b \\
c \qquad a=c \;\textrm{or}\; b=c
\label{eq:majority}
\end{cases}.
\end{align}
Since, hereafter, there appears no case that all variables are distinct, this definition is well-defined in our usage. In particular, we extend this to the case that the variables are pyramids as follows :
\begin{align}
mj(\hat{\lambda_A},\hat{\lambda_B},\hat{\lambda_C}):=\{mj((\hat{\lambda_A})_{h:\alpha},(\hat{\lambda_B})_{h:\alpha},(\hat{\lambda_C})_{h:\alpha})\}_{h:\alpha}.
\end{align}

Finally, we define $q$-dimension function $D$ :
\begin{align}
D[\hat{\lambda}]:=\prod_{h=1}^{N-1}\prod_{\alpha=1}^{N-h} \dfrac{[(\hat{\lambda})_{h:\alpha}+h]_q}{[h]_q}.
\label{eq:dimension of pyramid}
\end{align}

Now that we get all necessary tools, let us write down the local Boltzmann factors for three dominant weights $\lambda_A,\lambda_B$ and $\lambda_C$ living on its vertices.
This is expressed as
\begin{align}
B^{\triangle}_{\lambda_A,\lambda_B,\lambda_C}=\dfrac{1}{D[mj(\hat{\lambda_A},\hat{\lambda_B},\hat{\lambda_C})]^{\tfrac{1}{2}}}.
\label{eq:final exp. for local Boltzmann factor}
\end{align}
Based on this proposal, we derive several skein relations in App.~\ref{app:derivation of skeins}, which provides a (mathematical) evidence that this proposal works well. As we will see in the following examples, an implication of global symmetry enhancements also supports the validity of the above formula.

\subsection{Concrete computations}
\label{subsec:examples}

\begin{figure}[t]
\centering
\includegraphics[width=100mm]{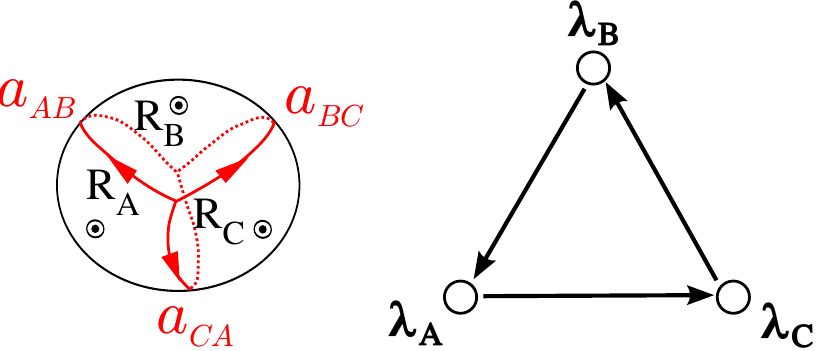}
\caption{Pants networks (Left) and their dual quivers (Right). In the dual system, there are two triangles forming a sphere.}
\label{fig:pants network}
\end{figure}

In this section, we see three examples : $T_3$-theory, $T_4$-theory including a Higgsed theory of it, and $T_{4 [N]}$-theories for general $N$.
$T_N$-theory is a SCFT in the case that $C$ is a two-sphere with three maximal punctures labelled by $[N]$. $T_{4 [N]}$-theory corresponds to a two-sphere with four maximal punctures.
In the first two cases, namely, $T_3$ and $T_4$-theories, these theories are Argyres-Seiberg dual theories to some theories and their Lagrangians are unknown yet \cite{ArgyresSeiberg07,MinahanNemeschansky96} although, for the $T_3$-theory, there is an interesting proposal that an $\SUSYN{1}$ Lagrangian theory flows in the IR to that theory \cite{GRW15}. There we consider elementary defects as shown in Fig.~\ref{fig:pants network}. They were discussed explicitly at first in \cite{Xie1304} and shown to be elementary generators of the line operator algebra in \cite{CGT15}. They are called pants networks there.
In the last example, we consider the two loops wrapping different one cycles as shown in Fig.~\ref{fig:dual loops}.
%\cite{GMN3,Neitzke1412}
%As long as we focus on the lowest order of $q$, it is possible to neglect the prefactor K(a)=1+\calO(q)$ and replace them by $1$.

\subsubsection*{$T_3$ theory}

In this theory, we can see that the above conjectural procedure exactly reproduces the previous result computed in \cite{TW1504}. There is just only pants network defect in $T_3$ theory up to charge conjugate operation, namely, $a_{AB}=a_{BC}=a_{CA}=1$.
Using the cyclic symmetry $A \to B \to C \to A$, we can take $E_{BC}=\{ 3 \}$ without loss of generality and write the Dynkin labels of $\lambda_C$ as $[n,m]$. Note that $\lambda_B=[n,m-1]$.
Then there are two sectors :
$(E_{AB},E_{CA})=(\{1\},\{2\})$ and $(\{2\},\{1\})$.
The former and latter give $\lambda_A=[n+1,m-1]$ and $[n-1,m]$, respectively, and the local Boltzmann factors are given by
\begin{align}
B^\triangle&=\dfrac{[2]}{[n+1][m][n+m+2]} \qquad \for ([n,m],[n,m-1],[n+1,m-1])\\
B^\triangle&=\dfrac{[2]}{[n+1][m+1][n+m+1]} \qquad \for ([n,m],[n,m-1],[n-1,m]).
\end{align}

\subsubsection*{$T_4$ theory and $T_{[4],[4],[2^2]}$ theory}

%At first, we see the configurations contributing to the lowest order of $q$ and then discuss the 

There is only $(2,1,1)$-junction and, therefore, there are three types of pants networks. However, their expectation values are exchanged under the cyclic permutation $a_{AB} \to a_{BC} \to a_{CA} \to a_{AB}$ and we can set $|E_{AB}|=2$ without loss of generality.
There are 12 distinct sectors for $(2,1,1)$-junction.
%At the leading order of $q$, The Schur index is given by
If we expand the $q$-deformed Yang-Mills expectation values in terms of $q$, this is given by
\begin{align}
\nonumber & q^{1/2} \chi_{\overline{\Box}}(a) \chi_{\Box}(b)
+q \left[\chi_{\tiny \yng(1,1)}(b)\chi_{\Box}(c)+\chi_{\tiny \yng(1,1)}(a)\chi_{\overline{\Box}}(c)\right]
+q^{3/2} \chi_{\Box}(a)\chi_{\overline{\Box}}(b) \chi_{\tiny  \yng(1,1)}(c) \\
+& q^2\left[\chi_{\rm Adj}(a)(\chi_{\tiny \overline{\yng(2)}}(b)+\chi_{\tiny \yng(1,1)}(b))\chi_{\Box}(c)
+(\chi_{\tiny  \yng(1,1)}(a)+\chi_{\tiny \overline{\yng(2)}}(a))\chi_{\rm Adj}(b)\chi_{\overline{\Box}}(c)
\right]+\calO(q^{5/2})
\end{align}
and the Schur index can be obtained by multiplying a prefactor $(q^2;q)_\infty(q^3;q)_\infty(q^4;q)_\infty$  $\displaystyle \prod_{x=a,b,c} \mathrm{P.E.}\left[ \dfrac{q}{1-q} \chi_{\rm Adj}(x)\right]$ where $(x;q)_\infty:=\prod_{i=0}^{\infty}(1-xq^i)$ and $\displaystyle \mathrm{P.E.}\left[ \dfrac{q}{1-q} \chi_{\rm Adj}(x)\right]:=\prod_{\alpha \in \Pi({\rm \mathbf{Adj}})}\dfrac{1}{(qx^\alpha;q)_\infty}$.
\!\footnote{For generic punctures, this prefactor is slightly changed because we must remove the NG modes associated with the symmetry breaking \cite{Tachikawa1504,GRRY1110,GaiottoRazamat12}.}

Next, let us consider $T_{[4],[4],[2^2]}$-theory where $C$ is a two-sphere with two maximal punctures and one $[2,2]$-type puncture.
The reason why we focus on this case is that this theory enjoys the global symmetry enhancement from the manifest global symmetry $SU(4) \times SU(4) \times SU(2)$ into $E_7$ global symmetry \cite{ArgyresSeiberg07,BBT09}.

On performing partially closing operations and the $q$-expansion, we must take account of the higher powers of $q$ in the above analysis because $\chi^{SU(4)}_R(x)$ is a series of $q^{1/2}$ including negative powers, where $x$ is a UV holonomy associated with the $[2,2]$-type puncture. See part.4 in Sec.~\ref{subsec:review on qYM} as to the partially closing operation.
By taking it into considerations, it turns out that there are following four possible configurations contributing to the lowest order of $q$ :
$(R_A,R_B,R_C)=\quad$ 1.$(\raisebox{-1ex}{\tiny \yng(1,1,1)},{\tiny \yng(1)},\phi)$,
2.$(\phi,\raisebox{-0.5ex}{\tiny \yng(1,1)},{\tiny \yng(1)})$,
3.$(\raisebox{-0.5ex}{\tiny \yng(1,1)},\phi,\raisebox{-1ex}{\tiny \yng(1,1,1)})$ and 
4.$({\tiny \yng(1)},\raisebox{-1ex}{\tiny \yng(1,1,1)},\raisebox{-0.5ex}{\tiny \yng(1,1)})$. In conclusion, we have
\begin{align}
I(a,b,c')&=q^{1/2}\;\chi^{E_7}_{\textbf{56}}(a,b,c')+\calO(q^{3/2}) \\
I(a',b,c)&=\chi^{SU(4)}_{\Box}(b)\chi^{SU(2)}_{\Box}(a')+\chi^{SU(4)}_{\ovl{\Box}}(c)+\calO(q) \\
I(a,b',c)&=\chi^{SU(4)}_{\ovl{\Box}}(a)\chi^{SU(2)}_{\Box}(b')+\chi^{SU(4)}_{\Box}(c)+\calO(q)
\end{align}
where $x'$ for $x=a,b$ and $c$ is related to $x$ by
$x=(q^{1/2}x',q^{-1/2}x',q^{1/2}x'^{-1},q^{-1/2}x'^{-1})$.
Recalling the fact that the Coulomb branch complex dimension of $T_{[4],[4],[2^2]}$ theory is $1$, this is consistent with this result that there is only one elementary network reflecting the global $E_7$ symmetry.

\begin{comment}
\subsubsection{$T_6$ theory and $T_{[6],[2^3],[3^2]}$ theory}

$E_8$ global symmetry

$T_6$ at first

$(4,1,1)$,$(3,2,1)$ and $(2,2,2)$-junctions

one of $(3,2,1)$ enjoys E8 symmetry

\subsubsection{$\SUSYN{2^\ast}$ gauge theory}

\end{comment}

\subsubsection*{Dual intersecting loops in $T_{4 {\rm fulls}}$}

\begin{figure}[t]
\centering
\includegraphics[width=40mm]{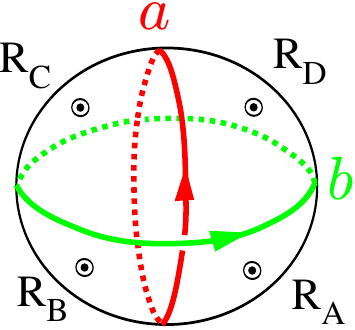}
\caption{Two dual intersecting loops. If the upper punctures are simple / minimum type $[2,1^{N-2}]$, the red loop corresponds to the fundamental Wilson loop and the green one does to some 't Hooft loop.}
\label{fig:dual loops}
\end{figure}

Let us consider the case with $a=b=1$ in Fig.~\ref{fig:dual loops}.
This theory reduces into $SU(N)$ superconformal QCD (SCQCD) on partially closing two of four punctures into the simple ($[2,1^{N-2}]$-type) punctures.
On that theory, these two loops correspond to the ordinary fundamental Wilson loop and some 't Hooft loop.
\footnote{At least, its 't Hooft's topological charge is neutral. It is an interesting problem to identify what line defect on the 4D SCQCD precisely corresponds to the given loops on the 2D geometry side.}
Using the crossing resolutions, this decomposes into four components. By evaluating each component and then by summing them up, we have the following results for the whole Boltzmann factor (the definition of $E_{XY}$ is given in Sec.~\ref{subsubsec:allowed}) :
\begin{enumerate}
\item case $E_{BA}=E_{DA}=E_{CB}=E_{CD}=\{ \ell \}$ for $\ell \in \{1,2,\ldots,N\}$
\begin{align}
B_{\beta \circ \alpha:\{\lambda\}}=\dfrac{1}{D[\hat{\lambda_B}]^2}
\end{align}
where $\lambda_B=\lambda_D$.

\item case $E_{BA}=E_{DA}=\{ \ell \}$ and $E_{CB}=E_{CD}=\{ k \}$ for $\ell \neq k \in \{1,2,\ldots,N\}$
\begin{align}
B_{\beta \circ \alpha:\{\lambda\}}=\dfrac{1}{D[\hat{\lambda_B}]^2[\kappa+\sigma_0]_q^2}
\end{align}
where $h_0:=|k-\ell|$, $\alpha_0:=\min(k,\ell)$, $\sigma_0:=\sgn(\ell-k)$ and $\kappa:=(\hat{\lambda_B})_{h_0:\alpha_0}+h_0$. Note also $\lambda_B=\lambda_D$.

\item case $E_{CD}=E_{BA}=\{ \ell \}$ and $E_{DA}=E_{CB}=\{ k \}$ for $\ell \neq k \in \{1,2,\ldots,N\}$
\begin{align}
B_{\beta \circ \alpha:\{\lambda\}}=\dfrac{1}{D[\hat{\lambda_B}+\hat{f}_{\{\ell\},\{k\}}]D[\hat{\lambda_D}+\hat{f}_{\{k\},\{\ell\}}]}
\end{align}
where $\lambda_B-h_k=\lambda_D-h_\ell$ and see App.~\ref{app:more math} as to $\hat{f}_{\{k\},\{\ell\}}$.

It is possible to rewrite the above expression into
\begin{align}
B_{\beta \circ \alpha:\{\lambda\}}=\dfrac{1}{D[\hat{\lambda_B}]D[\hat{\lambda_D}]}
\dfrac{[\kappa]_q[\kappa+2\sigma_0]_q}{[\kappa+\sigma_0]_q^2}.
\end{align}
There is a relation $\kappa=(\hat{\lambda_B})_{h_0:\alpha_0}+h_0=(\hat{\lambda_D})_{h_0:\alpha_0}+h_0-2\sigma_0$.

\end{enumerate}

Note that the ordering of additions of the two loops is irrelevant in $q$-deformed Yang-Mills theory (not so in the Liouville-Toda CFT case) and they commutes each other. We can naturally understand this if we put them in three dimensional space $C \times S^1$ as discussed in Sec.~\ref{sec:composite}.

%It is possible to unify the above three cases introducing a new function
%In particular, focusing on $N=2$ case, there $T_{4 {\rm full}}$-theory is exactly the SCQCD and we identify the representation with its dimension.

\subsection{A remark on $\calR$-matrix}
\label{subsec:remark on R-matrix}

\begin{figure}[th]
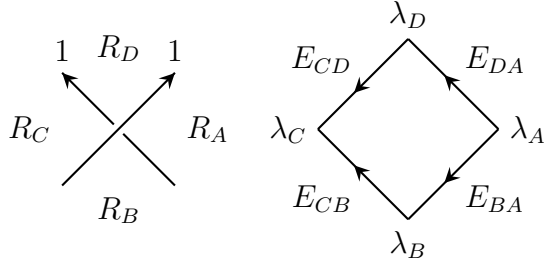

\centering
\begin{tabular}{cc}
\begin{minipage}{0.5\hsize}
\raggedleft
\input{figure/sec2/Rmatrix.tikz}
\end{minipage}
\begin{minipage}{0.5\hsize}
\raggedright
\input{figure/sec2/dualRmatrix.tikz}
\end{minipage}
\end{tabular}
\caption{There are four regions around any crossing. This means that the $\calR$-matrices (Left) can be mapped into the local Boltzmann factors associated with rectangles (Right).}
\label{fig:Rmatrix}
\end{figure}
As we see in the last example in the previous section,
it is possible to compute the local Boltzmann factor for a single crossing or on the dual rectangle. See Fig.~\ref{fig:Rmatrix}.
%The extension to the general crossing pairs is straightforward though we have no proof.
Roughly speaking, the factors are the square root of the previous results, but there appear some additional powers of $q$.
The factors can be given as follows :
\begin{enumerate}
\item case $E_{BA}=E_{DA}=E_{CB}=E_{CD}=\{ \ell \}$
\begin{align}
B^{\Box}_{\rm crossing}=(-q^{1/2})^{\tfrac{1}{N}}\dfrac{-q^{-1/2}}{D[\hat{\lambda_B}]}
\end{align}
where $\lambda_B=\lambda_D$ again.

\item case $E_{BA}=E_{DA}=\{ \ell \}$ and $E_{CB}=E_{CD}=\{ k \}$ for $\ell \neq k$
\begin{align}
B^{\Box}_{\rm crossing}=(-q^{1/2})^{\tfrac{1}{N}}\dfrac{-\sigma_0 q^{-\sigma_0 \kappa-1}}{D[\hat{\lambda_B}][\kappa+\sigma_0]_q}
\end{align}
where we use the same $\kappa$ and $\sigma_0$ as before. Note also $\lambda_B=\lambda_D$ again.

\item case $E_{CD}=E_{BA}=\{ \ell \}$ and $E_{DA}=E_{CB}=\{ k \}$ for $\ell \neq k$
\begin{align}
B^{\Box}_{\rm crossing}=(-q^{1/2})^{\tfrac{1}{N}}\dfrac{1}{D[\hat{\lambda_B}]^{1/2}D[\hat{\lambda_D}]^{1/2}}
\dfrac{[\kappa]_q^{1/2}[\kappa+2\sigma_0]_q^{1/2}}{[\kappa+\sigma_0]_q}.
\end{align}
\end{enumerate}
The other crossing is obtained by just replacing $q$ by $q^{-1}$.
All the above results can have similar structures to those for triangles.

It is a very interesting problem to analyze all types of crossings or to relate the above local Boltzmann factors to the known models such as face or (R)SOS models \cite{ABF84,FZ89,book:GAS,Pasquier86,Costello1308}.
%\footnote{Intuitively, majority function measure the step in SOS model...}

%% file: figure/sec2/right_arrow.tikz
\tikzsetnextfilename{-figure-sec2-right_arrow}
\begin{tikzpicture}
\def\l{1.6}
\coordinate (s) at (0,0);
\coordinate (t) at ($(s)+(\l,0)$);

\draw[end arrow] (s) -- (t) node[right]{$R$};
\end{tikzpicture}

%% file: figure/sec2/left_arrow.tikz
\tikzsetnextfilename{-figure-sec2-left_arrow}
\begin{tikzpicture}
\def\l{1.6}
\coordinate (s) at (0,0);
\coordinate (t) at ($(s)+(-\l,0)$);

\draw[end arrow] (s) -- (t) node[left]{$R^\ast$};
\end{tikzpicture}

%% file: figure/sec2/general_R.tikz
\tikzsetnextfilename{-figure-qYM-general_R}
\begin{tikzpicture}
\def\l{1.6}
\coordinate (UL) at (0,0);
\coordinate (DL) at ($(UL)+(0,-\l)$);
\coordinate (UR) at ($(UL)+(\l,0)$);
\coordinate (DR) at ($(UR)+(0,-\l)$);

\draw[end arrow] (DR) -- (UL) node[above]{$a$};
\draw[line width=5.0pt,white] (DL) -- (UR);
\draw[end arrow] (DL) -- (UR) node[above]{$b$};
\end{tikzpicture}

%% file: composite_defects.tex
\section{Composite surface-line systems}
\label{sec:composite}

%Although we do not specify geometries where the 4D gauge theories live,
%we suppose there two types of geometries : $S^4$ and $S^3 \times S^1$

As explained in the introduction, in the 2D system, the geometric counterparts of 4D line operators and 4D surface operators are networks and punctures, respectively.
As long as we treat either only surface operators or only line operators, the projection onto $C$ is natural to discuss the 4D physics.
However, if we have line defects bounded on 2D surface defects, it is not unique picture and there appears the new direction which line defects are localized in but surface defects extend along.

% % topological invariance + projection invariance

In Sec.~\ref{subsec:review_composite},
we review several basic facts needed later.
Next, we take a look at the geometrical configurations of two types of defects in Sec.~\ref{subsec:geometrical configurations}.
In Sec.~\ref{subsec:punctured skein relations},
we discuss the skein relations including fully degenerate punctures.
First, by introducing new topological moves, we derive such skein relations in some simple cases.
Then, in the last Sec.~\ref{subsec:other projection}, we rederive more general skein relations assuming the projection invariance.

\subsection{Brief review}
\label{subsec:review_composite}

\subsubsection*{Notation}

For a maximal torus element $a \in \bbT^{N-1}$ of $SU(N)$ and a weight vector $\lambda \in \Lambda_{wt} \simeq \bbZ^{N-1}$, we introduce a symbol
$a^\lambda:= (a_1^{\subul{\lambda}{1}},a_2^{\subul{\lambda}{2}},\ldots,a_N^{\subul{\lambda}{N}})$
where $\subul{\lambda}{i}:=( \lambda,h_i ) \quad i=1,2,\ldots,N$.
\footnote{Note that we keep the symbol $\lambda_\alpha$ as the Dynkin labels which are coefficients of $\omega_\alpha$ of the highest weight. See App.~\ref{app:more math} as for Lie algebra notations. The inner product $(,)$ is also defined there.}
$\rho$ denotes the Weyl vector, which is defined as $\displaystyle \sum_{a=1}^{N-1} \omega_a$.
$C_2(R)$ is the quadratic Casimir defined as $(\lambda,\lambda+2\rho)$ which is nomalized as $C_2(\Box)=N-\tfrac{1}{N}$.
We also introduce a symbol $\sigma_R=(-1)^{|R| (N-1)}$ where $|R|$ is the number of the boxes in the Young diagram corresponding to the irreducible representation $R$.

To avoid the appearance of numerical complex factors, through this section, we use the Liouville-Toda convention, which is the same as that used in Sec.3 in \cite{TW1504}.
See the footnote ~\ref{fn:convention} as to this point.

\subsubsection*{Skein relations}

We show two necessary skein relations in this section. See \cite{TW1504} for more relations.
One type is crossing resolutions already shown in the previous section.
\begin{eqnarray}
\raisebox{-0.4\height}{\input{figure/skein_relation/general_R.tikz}}
=
\frq^{\frac{ab}{N}}
\sum_{i=0}^{c}
\frq^{-i}
\raisebox{-0.5\height}{\input{figure/skein_relation/general_Q.tikz}}
\label{eq:crossing resolution}
\end{eqnarray}
where $c:=\min(a,b,N-a,N-b)$. Hereafter, using ~\eqref{eq:flip of arrow and cc of rep}, we only consider the case $a,b \le \tfrac{N}{2}$ for simplicity.
Note that each part of the right hand has a relation like
\begin{align}
\raisebox{-0.5\height}{\input{figure/skein_relation/general_Q.tikz}}
=\raisebox{-0.5\height}{\input{figure/skein_relation/general_Q2.tikz}}.
\end{align}

The other one is Reidemeister move I :
\begin{align}
\raisebox{-0.5\height}{\input{figure/skein_relation/RmI_1.tikz}}
=\sigma_R \frq^{-C_2(R)}\;
\raisebox{-0.5\height}{\input{figure/skein_relation/RmI_0.tikz}},
\qquad\qquad
\raisebox{-0.5\height}{\input{figure/skein_relation/RmI_2.tikz}}
=\sigma_R\frq^{C_2(R)}\;
\raisebox{-0.5\height}{\input{figure/skein_relation/RmI_0.tikz}}.
\label{eq:R-move I}
\end{align}
where $R$ can be any irreducible representation other than fundamental representations $\wedge^a \Box$.

\subsubsection*{Surface defect}

%First of all, note that there are two types of surface defect configurations.
It was discussed in \cite{GRR1207} that the SCIs in the presence of surface defects can be physically obtained by coupling a free hypermultiplet carrying $U(1)$ baryon symmetry to the original theory at UV and by taking the IR limit of that theory after giving variant VEVs to Higgs branch operators.
At the mathematical level, this corresponds to taking the residues at a pole in the fugacity complex planes, associated with the surface operator's charges and finally results in a difference operator acting on the flavor fugacities of the original theory.
%This operation means there are charged defects 

The above procedure is expected to reproduce in the IR the same defects as those from codimension four defects in 6D $\SUSYN{(2,0)}$ SCFTs and, in fact, this was checked in \cite{GaddeGukov13} by comparing these results with 4D SCIs coupled to the elliptic genera of the 2D $\SUSYN{(2,2)}$ theories living on the surface defects.
The difference operators in the Schur limit actually form the representation ring of $\Liesu(N)$ because the codimension four defects are labelled by representations of $\Liesu(N)$ \cite{ABFH13,BFHR14}.
%Since we include line defects, we consider one configuration only in this paper.
%They described the surface defects in terms of ...

According to \cite{GRR1207,ABFH13,BFHR14}, we rewrite the difference operator for the surface defect labelled by an irreducible representation $S$ as
\begin{align}
\widehat{\scrG}_{S}&=(\sqrt{I_{\textrm{vector}}(a)}) \cdot \left[\sum_{\lambda \in \Pi(S)} \frq^{-N ( \lambda,\lambda )} a^{N\lambda} \widehat{\Delta}_{-\lambda} \right]\cdot (\sqrt{I_{\textrm{vector}}(a)})^{-1} \\
&=(\sqrt{\Delta_{\rm Haar}(a)})^{-1}
\left[\sum_{\lambda \in \Pi(S)} \widehat{\Delta}_{-\lambda} \right]\cdot (\sqrt{\Delta_{\rm Haar}(a)})
%\prod_{i=1}^{N}a_i^{N ( \lambda,h_i )}
\label{eq:difference op. for surface defects}
\end{align}
where we have renormalized so that they form the representation ring of $\Liesu(N)$ exactly, $I_{\textrm{vector}}(a)$ is the SCI contribution from the vector multiplet whose concrete expression does not matter in this paper, $\Delta_{\rm Haar}(a)$ is the Haar measure of $SU(N)$ and $\widehat{\Delta}_{-\lambda}$ acts on a holonomy $a$ by $\frq^{-2\lambda} a$.
The characters $\chi_R(a)$ are common eigenfunctions of these operators for any $S$ and their eigenvalues are given by
\begin{align}
\bar{\calE}^{(S)}_R = \chi_{S}(\frq^{-2(\rho+\lambda_R)})=\dfrac{\dim_\frq S}{\dim_\frq R} \chi_{R}(\frq^{-2(\rho+\lambda_S)}).
\label{eq:eigenvalues of difference operators}
\end{align}

Finally, we remark on the mathematical relation between the codimension four defects and the codimension two defects ~\cite{DGG10}.
In the Liouville-Toda CFT set-up,
the general vertex operator is given by
$V_{\alpha}(z)=:e^{\langle \alpha,\phi(z) \rangle}:$
where $\alpha$ is a vector in $\GNOL{\Lieh}$ which is the dual to Cartan subalgebra, $z \in C$ and $\phi(z)$ is the Liouville-Toda scalar field.
This corresponds to the general codimension two defect ($[N]$ type or maximal puncture) when $\alpha-(b+1/b)\rho \in i\bbR^{N-1} \simeq \GNOL{\Lieh}$.
On the other hand, the codimension four defect labelled by a $\Liesu(N)$ irreducible representation $S$ is obtained by taking the limit $\alpha \to -b \lambda_S$ or $-\dfrac{1}{b} \lambda_S$.
\footnote{The two limits correspond to two types of configurations of surface defects in $S^4_b$. See the next subsection ~\ref{subsec:geometrical configurations}.}
The vertex operator in this limit is called fully degenerate and we also refer to the corresponding punctures as fully degenerate punctures which exactly represent the 4D surface defects.

In the 2D $q$-deformed Yang-Mills theory, the procedure similar to the above one is given as
\begin{align}
\lim_{a \to q^{-\rho-\lambda}} \dfrac{\chi_{R}(a)}{\dim_\frq R}
=\dfrac{\bar{\calE}^{(S)}_R}{\dim_\frq S}
\label{eq:codimension four from two}
\end{align}
where the denominator on the right hand side is just simply the normalization factor of the surface defect.
In our normalization, the surface defects exactly reproduce the $\Liesu(N)$ representation ring :
\begin{align}
\widehat{\scrG}_{S_1} \circ \widehat{\scrG}_{S_2}=\sum_{S_3} N_{S_1S_2}^{\qquad S_3}\;\widehat{\scrG}_{S_3}
\qquad \mbox{or} \qquad
\bar{\calE}^{(S_1)}_R \bar{\calE}^{(S_2)}_R =\sum_{S_3} N_{S_1S_2}^{\qquad S_3} \;\bar{\calE}^{(S_3)}_R.
\end{align}

\subsection{Geometrical configurations}
\label{subsec:geometrical configurations}

Originally, the bulk geometry of 6D $\SUSYN{(2,0)}$ SCFT is $S^3 \times S^1_E \times C$.
Both surface defects and line defects in 4D wrap $S^1_E$.
After the $S^1_E$ reduction, the geometry is the product of $C$ and a $S^1_H$-fibration over $S^2$ in our viewpoint.
$S^1_H$ is a Hopf fiber which is the support of surface defects in 4D.
\footnote{There are at least two kinds of surface defects when line defects are absent. The other one is obtained by exchanging two SCI fugacities $p$ and $q$ as seen in \cite{GRR1207}.}
On the other hand, line defects in 4D are networks on $C$.
Therefore, both types of defects are knots with junction in the fiber geometry $S^1_H \times C =: M$ and localized at the same point in base geometry $S^2$.
In the following discussion, we regard $S^1_H$ as an interval $I_H$ whose two end points are identified.
Then let $C_{in}$ and $C_{out}$ denote two boundaries of $I_H \times C$.
We interpret the surface defects as a defect running from a point in $C_{in}$ to the same point in $C_{out}$ along the $I_H$-direction.

%In particular, the 4D-2D-1D data are given by the (ambient) isotropy class.
%\footnote{Here we do not pay attentions to the boundary positions.}

Note that if we consider a 6D $\SUSYN{(2,0)}$ SCFT on $S^4 \times C$, a similar argument holds true. This is because OPEs of two BPS defects are expected to be determined locally and independent from the global background geometry.
Concretely, a surface defect extends along a $S^2=\{ (z,w=0,x) \in \bbC \times \bbC \times \bbR \;|\; b^2|z|^2 + x^5 = 1 \}$ in $S^4=\{ (z,w,x) \in \bbC \times \bbC \times \bbR \;|\; b^2|z|^2 + b^{-2} |w|^2 + x^5 =1 \}$ \cite{GomisLeFloch14} and some line defects live on $S^1=\{ (z,w=0,x_5=x_\ast) \in \bbC \times \bbC \times \bbR \}$ where $x_\ast$ is an arbitrary constant satisfying $|x_\ast|<1$ \cite{HamaHosomichi1209}.
Since only the local geometry around the defect locus is relevant, instead of $S^1_H$, we take the new direction as the $x^5$ direction (open interval) in this case. Therefore, it is expected that the skein relations discussed in Sec.~\ref{subsec:punctured skein relations} are also applied to the Liouville-Toda CFTs and we can check, in several examples, the claim that they are common in both systems. The relation between $\frq$ and $b$ is given in \cite{DGG1112} or \cite{TW1504} as $\frq=e^{i\pi b^2}$.

The phenomenon inherent in the $S^4_b$ case is that there simultaneously exist two types of line operators and, in a such case, it seems to be necessary to treat them in the full five dimensional geometry rather than three dimensional one.
Notice also that there are two distinct origins of the non-commutativity of line operators correspondingly. One comes from the Poynting vector in the bulk generated by line's charges as discussed in \cite{GMN3,IOT11} and this classical picture also may be valid in the Schur index case. The other interpretation is similar but different. There, both line operators cannot be genuine line operators and either should be the boundary of an open surface operator. Then, two operators have some contact interactions under the exchange of their ordering in the 4D bulk \cite{AST13,KapustinSeiberg14,GukovKapustin13}.

\subsection{Skein relations with fully degenerate punctures}
\label{subsec:punctured skein relations}

% % % put the following sentence somewhere
%same as the label of edges in networks ... we restrict the representation -> surface operator into the fundamental representations
%they generate all the irreducible representations if we allow the product (OPE) and addition (superpositions)

At first, we use the same projection of $M$ onto the 2D plane as before. This is the projection onto $C$ which we call ``$C$-projection".

If the 4D surface defects are topological in $M$, by deforming its orbit in $M$, we expect the following relation :
\begin{align}
\raisebox{-0.2\height}{\input{figure/skein_relation/puncture.tikz}}
=
\raisebox{-0.25\height}{\input{figure/skein_relation/puncture_with_framed_edge.tikz}}
=
\sigma_S\frq^{C_2(S)}\raisebox{-0.5\height}{\input{figure/skein_relation/puncture_with_edge1.tikz}}
=
\sigma_S\frq^{-C_2(S)}\raisebox{-0.4\height}{\input{figure/skein_relation/puncture_with_edge2.tikz}}.
\end{align}
Here we must take the framing factor appearing in R-move I ~\eqref{eq:R-move I} into consideration.

Let a white dot (a circle) and a black one (a filled circle) in the $C$-projection plane represent each intersection point of a surface defect with $C_{in}$ and $C_{out}$, respectively.
Then new moves appear :
\begin{align}
\raisebox{-0.35\height}{\input{figure/skein_relation/RmoveV1left.tikz}}
=
\raisebox{-0.183\height}{\input{figure/skein_relation/RmoveV1right.tikz}}
\qquad
\raisebox{-0.35\height}{\input{figure/skein_relation/RmoveV2left.tikz}}
=
\raisebox{-0.183\height}{\input{figure/skein_relation/RmoveV2right.tikz}}.
\end{align}
On the left hand side, a line labelled by $S$ stems from the white dot in $M$ and, on the right hand side, a line by $S$ goes into the black dot in $M$.
We call this relation Reidemeister move V (R-move V).
In particular, because two types of dots are identified in $S^1_H$, they coincide in $C$-projection and we have
\begin{align}
\raisebox{-0.45\height}{\input{figure/skein_relation/RmoveV_combined_left.tikz}}
=
\raisebox{-0.45\height}{\input{figure/skein_relation/RmoveV_combined_right.tikz}}.
\label{eq:RmoveV_combined}
\end{align}
We refer to the edges with dots on it as ``punctured edges".
Be aware that the punctured edges are just open lines in the three dimensional space $M$.
A view from the right hand towards the left hand is shown in Fig.~\ref{fig:other view for punctured edge}. See also Sec.~\ref{subsec:other projection} for the detail.
\begin{figure}[ht]
\centering
\begin{align}
\raisebox{-0.45\height}{\input{figure/skein_relation/punctured_edge.tikz}}
\longleftrightarrow
\raisebox{-0.45\height}{\input{figure/skein_relation/other_view_for_punctured_edge.tikz}}
\end{align}
\caption{A punctured edge labelled by $S$ in the left can be depicted as the right in the projection from $M$ onto other 2D plane extending along the Hopf fiber direction.}
\label{fig:other view for punctured edge}
\end{figure}

What we are interested in is the situation where a line in $C$ passes near a fully degenerate puncture.
The above relation ~\eqref{eq:RmoveV_combined} leads to
\begin{align}
\raisebox{-0.45\height}{\input{figure/skein_relation/RmoveV_application_left.tikz}}
=
\raisebox{-0.45\height}{\input{figure/skein_relation/RmoveV_application_right.tikz}}
\end{align}
and now we can apply the crossing resolutions ~\eqref{eq:crossing resolution} to the network representation on the right hand.
\footnote{Another more useful way to derive the same result is to separate the locations of ingoing and outgoing punctures (white and black dots) in $C$ firstly, to apply the skein relations secondly and to merge them again finally.}
%This relation tells us the OPE of a surface defect and a line defect in the geometrical language as we see.

\subsubsection*{Special case}

Let us take $S$ and $R$ as $\Box$ (or $1$) and $\wedge^{k} \Box$ (or $k$), respectively.
% % we abbreviate 1,k
There are two ways to do the calculations :
\begin{align}
\raisebox{-0.4\height}{\input{figure/skein_relation/ex1st_1.tikz}}
=
\sigma_{\Box}\frq^{C_2(\Box)}
\raisebox{-0.4\height}{\input{figure/skein_relation/ex1st_2.tikz}}
=
\frq^{\tfrac{2k}{N}}
\raisebox{-0.4\height}{\input{figure/skein_relation/ex1st_3.tikz}}
+
(-1)^k \frq^{\tfrac{2k-1}{N}+k-1} (\frq - \frq^{-1})
\raisebox{-0.5\height}{\input{figure/skein_relation/ex1st_4.tikz}}.
\end{align}
On the other hand,
\begin{align}
\raisebox{-0.4\height}{\input{figure/skein_relation/ex1st_1.tikz}}
=
\sigma_{\Box}\frq^{-C_2(\Box)}
\raisebox{-0.4\height}{\input{figure/skein_relation/ex1st_5.tikz}}
=
\frq^{\tfrac{2k}{N}-2}
\raisebox{-0.4\height}{\input{figure/skein_relation/ex1st_3.tikz}}
+
(-1)^{k-N-1} \frq^{\tfrac{2k+1}{N}+k-1-N} (\frq - \frq^{-1})
\raisebox{-0.5\height}{\input{figure/skein_relation/ex1st_6.tikz}}.
\end{align}
Comparing both expressions, we have
\begin{align}
\raisebox{-0.4\height}{\input{figure/skein_relation/ex1st_3.tikz}}
&=
(-1)^{k-N-1} \frq^{\tfrac{1}{N}+k-N}
\raisebox{-0.5\height}{\input{figure/skein_relation/ex1st_6.tikz}}
+
(-1)^{k+1}\frq^{k-\tfrac{1}{N}}
\raisebox{-0.5\height}{\input{figure/skein_relation/ex1st_4.tikz}} \\
&=
\frq^{-\tfrac{k}{N}}\left[
\raisebox{-0.45\height}{\input{figure/skein_relation/ex1st_8.tikz}}
+
\frq
\raisebox{-0.45\height}{\input{figure/skein_relation/ex1st_7.tikz}}
\right].
\label{eq:punctured skein relation (k,1)}
\end{align}

In the same way, we also have
\begin{align}
\raisebox{-0.4\height}{\input{figure/skein_relation/ex1st_1.tikz}}
=
\frq^{\tfrac{k}{N}}\left[
\raisebox{-0.45\height}{\input{figure/skein_relation/ex1st_8.tikz}}
+
\frq^{-1}
\raisebox{-0.45\height}{\input{figure/skein_relation/ex1st_7.tikz}}
\right].
\label{eq:punctured skein relation (1,k)}
\end{align}

%The representative elements are not unique.
%\begin{align}
%\raisebox{-0.4\height}{\input{figure/skein_relation/ex1st_8.tikz}}
%=
%\raisebox{-0.4\height}{\input{figure/skein_relation/ex1st_9.tikz}}
%\end{align}

In the simplest case $N=2$ and $k=1$ which do not need any junctions and arrows on edges, this becomes simpler as follows :
\begin{align}
\raisebox{-0.4\height}{\input{figure/skein_relation/simplest_exR.tikz}}
&=
\frq^{-\tfrac{1}{2}}
\raisebox{-0.45\height}{\input{figure/skein_relation/simplest_ex1.tikz}}
+
\frq^{\tfrac{1}{2}}
\raisebox{-0.45\height}{\input{figure/skein_relation/simplest_ex2.tikz}}
\label{eq:punctured skein relation (1,1) A1 case}
\end{align}
and the other relation can be obtained by mapping $\frq$ to $\frq^{-1}$.

If we apply either relation to the loop wrapping a cylinder and one fully degenerate puncture near it, there appear two kinds of knots.
One winds around the cylinder by one time as it goes from $C_{in}$ to $C_{out}$ and the other does in the opposite way.
Recalling the fact that there lives a 2D $\SUSYN{(2,2)}$ $U(1)$ gauged linear $\sigma$ model on the surface defect labelled by $\Box$ \cite{GaddeGukov13,GomisLeFloch14}, it is expected that these loops in $M$ represent the $U(1)$ Wilson loops charged $\pm 1$ according to the widing orientation, in the 2D system on the surface defect.

\subsubsection*{General case}

How do the similar relations look like for any pair $S=\wedge^{\ell}\Box$ and $R=\wedge^k \Box$ ?
From the above examples, we can expect that the general skein relations are
\begin{align}
\raisebox{-0.45\height}{\input{figure/skein_relation/OPE_network_puncture.tikz}}
=
\frq^{\tfrac{k\ell}{N}}\sum_{s=0}^{\min(k,\ell)} \frq^{-s}
\raisebox{-0.45\height}{\input{figure/skein_relation/OPE_network_puncture_component.tikz}}
\label{eq:OPE of network and puncture}
\end{align}
where the coefficients are the same as those of the crossing resolution \eqref{eq:crossing resolution}. The other relations are
\begin{align}
\raisebox{-0.45\height}{\input{figure/skein_relation/OPE_network_puncture2.tikz}}
=
\frq^{-\tfrac{k\ell}{N}}\sum_{s=0}^{\min(k,\ell)} \frq^{s}
\raisebox{-0.45\height}{\input{figure/skein_relation/OPE_network_puncture_component.tikz}}.
\label{eq:OPE of network and puncture2}
\end{align}
In the case of $\ell=1$, each reduces into ~\eqref{eq:punctured skein relation (k,1)}
or ~\eqref{eq:punctured skein relation (1,k)}.
We see in the next subsection ~\ref{subsec:other projection} that these relations are indeed reproduced in another approach.

\subsection{Other projections}
\label{subsec:other projection}

The requirement of the topological property of networks in $M$ means that their projection onto a 2D plane can be taken arbitrarily.
So far we have used $C$-projection, but actually, we can consider other projections onto a plane extending along $I_H$-direction. We call those projections ``$H$-projections".

Now it is possible to directly obtain the same result as before by applying the skein relation in a $H$-projection. Let us view the crossing network on the left hand side in ~\eqref{eq:OPE of network and puncture2} from the right hand and apply the crossing resolution on the new projection. This can be expressed as
\begin{align}
\raisebox{-0.4\height}{\input{figure/skein_relation/other_projection_cross.tikz}}
=
\frq^{\tfrac{k\ell}{N}}\sum_{s=0}^{\min(k,\ell)} \frq^{-s}
\raisebox{-0.5\height}{\input{figure/skein_relation/other_projection_resolved.tikz}}.
\end{align}
This relation exactly matches with the previous expressions ~\eqref{eq:OPE of network and puncture2} and we have a relation between distinct projections like
\begin{align}
\raisebox{-0.45\height}{\input{figure/skein_relation/OPE_network_puncture_component.tikz}}
=
\raisebox{-0.5\height}{\input{figure/skein_relation/other_projection_resolved.tikz}}
\end{align}
where the left hand side is the usual $C$-projection but the right one is a $H$-projection including $I_H$ direction.

%\subsubsection*{A few comments}
Finally, we make a brief comment on the reproduction of the relation ~\eqref{eq:codimension four from two}.
This can be geometrically expressed as
\begin{align}
\raisebox{-0.5\height}{\input{figure/applied/loop_winding_puncture_onto_C.tikz}}
=
\sigma_R \chi_R(q^{-\rho-\lambda_S})
\raisebox{-0.5\height}{\input{figure/applied/puncture_onto_C.tikz}}
\end{align}
or locally
\begin{align}
\raisebox{-0.5\height}{\input{figure/applied/loop_winding_puncture.tikz}}
=
\sigma_R \chi_R(q^{-\rho-\lambda_S})
\raisebox{-0.5\height}{\input{figure/applied/puncture.tikz}}.
\end{align}
We can derive this relation in some simple cases.

%$S=\wedge^k \Box$ and $M=\wedge^\ell \Box$
%$\Delta_{-\mu} a^{\lambda}=\frq^{}$
%$(-1)^{|R|(N-1)}\chi_{R}(q^{-\rho-\lambda_S})$

%% file: figure/skein_relation/general_R.tikz
\tikzsetnextfilename{-figure-skein_relation-general_R}
\begin{tikzpicture}
\def\l{1.6}
\coordinate (UL) at (0,0);
\coordinate (DL) at ($(UL)+(0,-\l)$);
\coordinate (UR) at ($(UL)+(\l,0)$);
\coordinate (DR) at ($(UR)+(0,-\l)$);

\draw[end arrow] (DR) -- (UL) node[above]{$a$};
\draw[line width=5.0pt,white] (DL) -- (UR);
\draw[end arrow] (DL) -- (UR) node[above]{$b$};
\end{tikzpicture}

%% file: figure/skein_relation/RmI_0.tikz
\tikzsetnextfilename{-figure-skein_relation-RmI_0}
\begin{tikzpicture}
\def\h{1.8}
\coordinate (U) at (0,0);
\coordinate (D) at ($(U)+(0,-\h)$);

\draw[end arrow] (D) node[below]{$R$} -- (U);
\end{tikzpicture}

%% file: figure/skein_relation/puncture.tikz
\tikzsetnextfilename{-figure-skein_relation-puncture}
\begin{tikzpicture}
\def\r{0.08}
\coordinate (P) at (0,0);

\draw[black,thick] (P) circle (1.8*\r) node[left]{$S \;$};
\draw[fill=black] (P) circle (\r);
\end{tikzpicture}

%% file: figure/skein_relation/RmoveV1left.tikz
\tikzsetnextfilename{-figure-skein_relation-RmoveV1left}
\begin{tikzpicture}
\def\l{1.0}
\coordinate (P) at (0,0);
\coordinate (U) at ($(P)+(0,\l)$);
\coordinate (L) at ($(P)+(200:\l)$);
\coordinate (R) at ($(P)+(-20:\l)$);

\draw[mid-end arrow] (P) -- (U) node[above]{$S$};
\draw[fill=white] (P) circle (0.08);
\draw[mid-end arrow] (L) -- (R) node[right]{$R$};
\end{tikzpicture}

%% file: figure/skein_relation/RmoveV2left.tikz
\tikzsetnextfilename{-figure-skein_relation-RmoveV2left}
\begin{tikzpicture}
\def\l{1.0}
\coordinate (P) at (0,0);
\coordinate (U) at ($(P)+(0,\l)$);
\coordinate (L) at ($(P)+(200:\l)$);
\coordinate (R) at ($(P)+(-20:\l)$);

\draw[start-mid arrow] (U) node[above]{$S$} -- (P);
\draw[fill=black] (P) circle (0.08);
\draw[mid-end arrow] (L) -- (R) node[right]{$R$};
\end{tikzpicture}

%% file: figure/skein_relation/simplest_ex1.tikz
\tikzsetnextfilename{-figure-skein_relation-simplest_ex1}
\begin{tikzpicture}
\def\l{1.0}
\def\r{0.08}
\coordinate (P) at (0,0);
\coordinate (U) at ($(P)+(0,\l)$);
\coordinate (D) at ($(P)+(0,-\l)$);

\draw[thick] (P) -- (U) node[above]{$1$};
\draw[fill=white,thick] (P) circle (1.8*\r);
\draw[fill=black] (P) circle (\r);
\draw[thick] (D) node[below]{$1$} -- (P);
\end{tikzpicture}

%% file: figure/skein_relation/simplest_ex2.tikz
\tikzsetnextfilename{-figure-skein_relation-simplest_ex2}
\begin{tikzpicture}
\def\l{1.0}
\def\r{0.08}
\coordinate (P) at (0,0);
\coordinate (U) at ($(P)+(0,\l)$);
\coordinate (D) at ($(P)+(0,-\l)$);

\draw[thick] (D) node[below]{$1$} -- (P);
\draw[fill=white,thick] (P) circle (1.8*\r);
\draw[fill=black] (P) circle (\r);
\draw[thick] (P) -- (U) node[above]{$1$};
\end{tikzpicture}

%% file: figure/applied/loop_winding_puncture_onto_C.tikz
\tikzsetnextfilename{-figure-applied-loop_winding_puncture_onto_C}
\begin{tikzpicture}
\def\r{0.08}
\def\rad{0.8}
\def\l{1.0}
\coordinate (C) at (0,0);
\coordinate (P) at ($(C)+(0:\rad)$);

\draw[fill=white] (C) circle (1.8*\r) node[left]{$S \;$};
\draw[black,thick] (C) circle (1.8*\r);
\draw[fill=black] (C) circle (\r);

\draw[thick] (C) circle (\rad);
\circlearrow{(C)}{\rad}{180}{0.3}{left}{\small $R$}
\end{tikzpicture}

%% file: figure/applied/puncture_onto_C.tikz
\tikzsetnextfilename{-figure-applied-puncture_onto_C}
\begin{tikzpicture}
\def\r{0.08}
\def\rad{0.45}
\def\l{1.0}
\coordinate (C) at (0,0);
\coordinate (P) at ($(C)+(0:\rad)$);

\draw[fill=white] (P) circle (1.8*\r) node[left]{$S \;$};
\draw[black,thick] (P) circle (1.8*\r);
\draw[fill=black] (P) circle (\r);
\end{tikzpicture}

%% file: figure/applied/loop_winding_puncture.tikz
\tikzsetnextfilename{-figure-applied-loop_winding_puncture}
\begin{tikzpicture}
\def\r{0.08}
\def\rad{0.45}
\def\l{1.0}
\coordinate (C) at (0,0);
\coordinate (P) at ($(C)+(0:\rad)$);
\coordinate (Q) at ($(C)+(180:\rad)$);
\coordinate (U) at ($(C)+(0,\l)$);
\coordinate (D) at ($(C)+(0,-\l)$);

\draw[thick] (P) arc (0:180:\rad);
\draw[line width=5.0pt,white] (C) -- (U);
\draw[end arrow] (C)  -- (U);
\draw[thick] (D) node[below]{$S$} -- (C);
\draw[line width=5.0pt,white] (Q) arc (180:300:\rad);
\draw[thick] (Q) arc (180:360:\rad);
\circlearrow{(C)}{\rad}{180}{0.3}{left}{\small $R$}
\end{tikzpicture}

%% file: figure/applied/puncture.tikz
\tikzsetnextfilename{-figure-applied-puncture}
\begin{tikzpicture}
\def\l{1.0}
\coordinate (C) at (0,0);
\coordinate (U) at ($(C)+(0,\l)$);
\coordinate (D) at ($(C)+(0,-\l)$);

\draw[end arrow] (D) node[below]{$S$} -- (U);
\end{tikzpicture}

%% file: open_network.tex
\section{Proposal for punctured network defects}
\label{sec:punctured network}
%define the terminology of punctured edge

We have independently discussed the computations of expectation values for closed network and the geometrical structures of the composite surface-line systems and here we will unify two things.
In Sec.~\ref{subsec:coexistence}, we compare the previous new skein relations with the computation of $q$-deformed Yang-Mills expectation values or the Schur indices.
From the comparison, we can extract the operator action of some punctured networks.
Based on this discussion, in Sec.~\ref{subsec:modified formula}, we propose the modified formula for general punctured networks and interpret the modification as the addition of the local Boltzmann factors assigned with dual arrowed edges.

\subsection{Coexistence of closed network and isolated punctures}
\label{subsec:coexistence}

In this section, let $C$ be a two-sphere with several punctures and $\gamma$ be a 2D Wilson loop wrapping a tube in $C$. This is the same situation as discussed in part 3 in Sec.~\ref{subsec:review on qYM}.
The general set-up can be discussed in the similar way.
Recalling the discussion around ~\eqref{eq:loop expression in q-YM}, let us cut along $\gamma$ and decompose the Riemann surface $C$ into the two parts which we call $C_{A}$ and $C_{B}$ here.
In the following, we see the operator structure in two distinct basis.

\subsubsection*{Fugacity/Holonomy basis}

The formula ~\eqref{eq:loop expression in q-YM} says that
the whole partition function is given by
\begin{align}
\oint[da] I_{C_A}(a,\ldots) \chi_M(a) I_{C_B}(a^{-1},\ldots).
\end{align}

\begin{figure}[ht]
\centering
\begin{tabular}{cc}
\begin{minipage}{0.5\hsize}
\centering
\includegraphics[width=70mm]{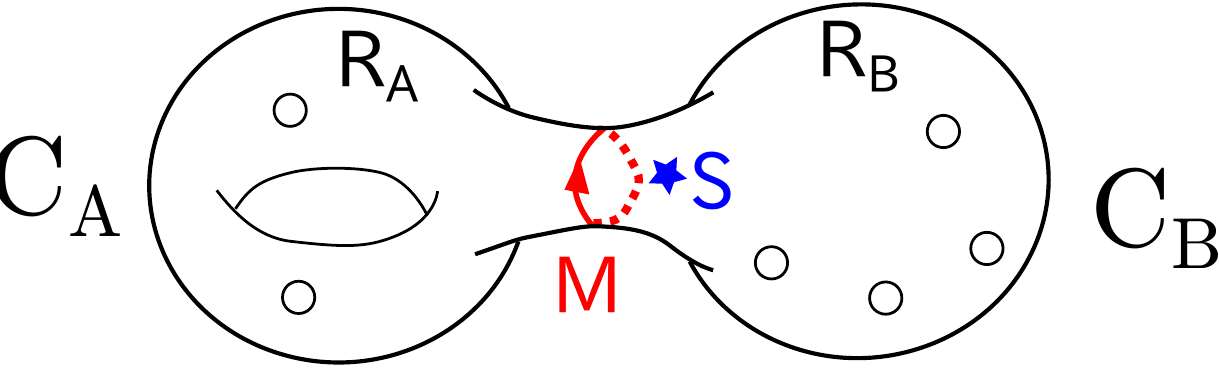}\\
\end{minipage}
\begin{minipage}{0.5\hsize}
\centering
\includegraphics[width=70mm]{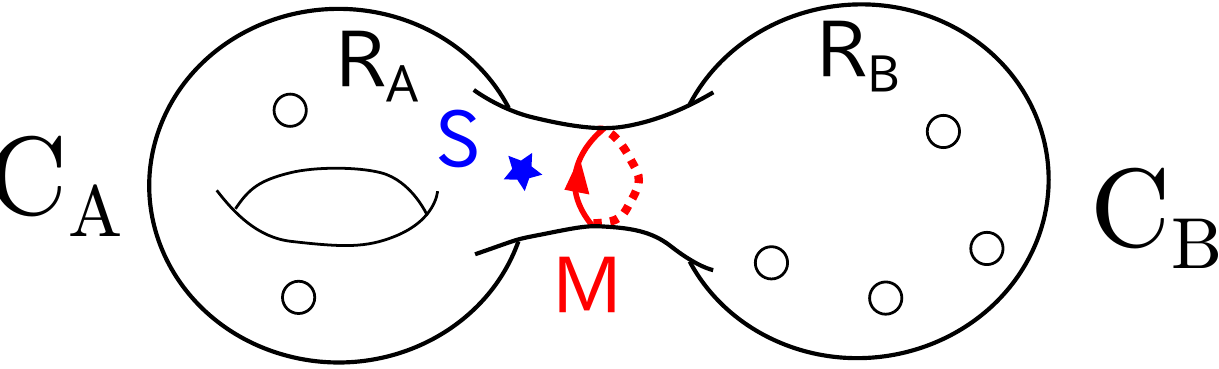}\\
\end{minipage}
\end{tabular}
\caption{Coexistence of a 2D Wilson loop and a fully degenerate puncture in $C$. Left corresponds to ~\eqref{eq:coexist_surface_line_1}~\eqref{eq:coexist_surface_line_1_rep_basis} and right does to 
~\eqref{eq:coexist_surface_line_2}~\eqref{eq:coexist_surface_line_1_rep_basis}.}
\label{fig:two choice of surface defect addition}
\end{figure}

Now let us add a surface defect labelled by $S$.
There are two choices of its addition, namely, the fully degenerate puncture on $C_A$ or on $C_B$ as shown in Fig.~\ref{fig:two choice of surface defect addition}.

They are evaluated as
\begin{align}
I_{C_A \sqcup_{W_M(\gamma)} (C_B,S)}:=
\oint[da] I_{C_A}(a,\ldots) \chi_M(a) (\widehat{\scrS}_{S}I_{C_B})(a^{-1},\ldots)
\label{eq:coexist_surface_line_1}
\end{align}
and
\begin{align}
I_{(C_A,S) \sqcup_{W_M(\gamma)} C_B}:=
&\oint[da] (\widehat{\scrS}_{S}I_{C_A})(a,\ldots) \chi_M(a) I_{C_B}(a^{-1},\ldots) \\
&=\oint[da] I_{C_A}(a,\ldots) \widehat{\scrS}_{S} (\chi_M(a) I_{C_B})(a^{-1},\ldots)
\label{eq:coexist_surface_line_2}
\end{align}
where we use the self-adjoint property of the difference operator $\widehat{\scrS}_{S}$.
And two expressions give different answers.

In particular, the special case $M=\wedge^\ell \Box$ and $S=\wedge^k \Box$ is important.
Let $\Pi(R)$ be the set of weights for an irreducible representation $R$ and we also introduce a subset defined as
\begin{align}
\Pi(\wedge^k \Box, \wedge^\ell \Box)_s:=\left\{
(\lambda,\mu) \in\Pi(\wedge^k \Box) \times \Pi(\wedge^\ell \Box)\;|\;
(\lambda,\mu)=s-\tfrac{k\ell}{N}
\right\}.
\end{align}
In the following network representations in this subsection, we identify two end points of any open edge in networks such that they once wrap a tube in $C$.

%Note :The calculation here is the same as those in 4.2 in %\cite{TW1504}
On one side, we have a relation like
\begin{align}
\raisebox{-0.45\height}{\input{figure/skein_relation/OPE_network_puncture2.tikz}}
\longleftrightarrow
\hat{W}_{\wedge^k \Box} \hat{\scrS}_{\wedge^\ell \Box}
=\sum_{\substack{\lambda \in \Pi(\wedge^k \Box) \\ \mu \in \Pi(\wedge^\ell \Box)}} a^{\lambda} \widehat{\Delta}^{\chi}_{-\mu}
=\sum_{s=0}^{\min(k,\ell)} \frq^{s-\tfrac{k\ell}{N}} \hat{\scrO}_{s}^{(k,\ell)}
\end{align}
where we have defined new difference operators conjugate to $\widehat{\Delta}_{-\lambda}$
\begin{align}
\widehat{\Delta}^{\chi}_{-\lambda}:=(\sqrt{\Delta_{\rm Haar}(a)})^{-1}\cdot \widehat{\Delta}_{-\lambda} \cdot (\sqrt{\Delta_{\rm Haar}(a)})
\end{align}
and another difference operator
\begin{align}
\hat{\scrO}_{s}^{(k,\ell)}
:=\sum_{ (\lambda,\mu) \in \Pi(\wedge^k \Box, \wedge^\ell \Box)_s }
a^{\lambda/2} \Delta^{\chi}_{-\mu} a^{\lambda/2}
=\sum_{ (\lambda,\mu) \in \Pi(\wedge^k \Box, \wedge^\ell \Box)_s }
\frq^{\tfrac{k\ell}{N}-s} a^{\lambda} \Delta^{\chi}_{-\mu}.
\end{align}
We also use the formula
\begin{align}
a^\lambda \widehat{\Delta}_{-\mu}=\frq^{2(\lambda,\mu)} \widehat{\Delta}_{-\mu} a^\lambda \qquad\quad
a^\lambda \widehat{\Delta}^{\chi}_{-\mu}=\frq^{2(\lambda,\mu)} \widehat{\Delta}^{\chi}_{-\mu} a^\lambda.
\end{align}

On the other hand, we have
\begin{align}
\raisebox{-0.45\height}{\input{figure/skein_relation/OPE_network_puncture.tikz}}
\longleftrightarrow
\hat{\scrS}_{\wedge^\ell \Box} \hat{W}_{\wedge^k \Box}
=\sum_{\substack{\lambda \in \Pi(\wedge^k \Box) \\ \mu \in \Pi(\wedge^\ell \Box)}} \widehat{\Delta}^{\chi}_{-\mu} a^{\lambda}
=\sum_{s=0}^{\min(k,\ell)} \frq^{\tfrac{k\ell}{N}-s} \hat{\scrO}_{s}^{(k,\ell)}.
\end{align}

Comparing \eqref{eq:OPE of network and puncture} and \eqref{eq:OPE of network and puncture2} with these results,
we naturally get the correspondence
\begin{align}
\raisebox{-0.45\height}{\input{figure/skein_relation/OPE_network_puncture_component.tikz}} \longleftrightarrow \hat{\scrO}_{s}^{(k,\ell)}.
\end{align}

In the special case $s=k=\ell$, we have
\begin{align}
\raisebox{-0.45\height}{\input{figure/open/punctured_loop.tikz}} \longleftrightarrow \hat{\scrO}_{k}^{(k,k)}=\frq^{-\tfrac{1}{N} k(N-k)}
\sum_{\lambda \in \Pi(\wedge^k \Box)} a^{\lambda}\widehat{\Delta}^{\chi}_{-\lambda}.
\end{align}

\subsubsection*{Representation basis}

We repeat the same analysis in another new basis.
For that purpose, let us expand the partition functions on $C_A$ and $C_B$ by the $SU(N)$ characters as
\begin{align}
& \calF_{R_A}(\{b\}) := \oint [da'] \chi_{R_A}(a'^{-1})I_{C_A}(a',\{b\}) \\
& \calG_{R_B}(\{c\}) := \oint [da']
\chi_{R_B}(a'^{-1}) I_{C_B}(a',\{b\})
\end{align}
and then we can express the expectation value of the Wilson loop in the representation $M$ as
\begin{align}
\langle \calF | \hat{W}_M | \calG \rangle
:=\sum_{R_{A},R_{B}} \oint[da] \chi_{R_{A}}(a^{-1})\calF_{R_A}(\{b\}) \chi_M(a) \chi_{R_{B}}(a)\calG_{R_B}(\{c\}).
\end{align}
where we introduced a matrix representation like
\begin{align}
& |\calF \rangle = \sum_R \calF_{R}(\{b\}) |R \rangle \\
& |\calG \rangle = \sum_R \calG_{R}(\{c\}) |R \rangle \\
& \langle R_1 | R_2 \rangle =\delta_{R_1,R_2} \qquad \mbox{orthonormal basis}.
\end{align}

Using the eigenvalues of difference operators ~\eqref{eq:eigenvalues of difference operators}, the addition of surface operators in this basis corresponding to ~\eqref{eq:coexist_surface_line_1} and ~\eqref{eq:coexist_surface_line_2} are expressed as
\begin{align}
\langle \calF | \hat{W}_M \hat{\scrS}_{S} | \calG \rangle
=\sum_{R_{A},R_{B}} N_{R_{B}S}^{\quad R_{A}} \calF_{R_A}(\{b\})
\bar{\calE}^{(S)}_{R_B} \calG_{R_B}(\{c\})
\label{eq:coexist_surface_line_1_rep_basis}
\end{align}
and
\begin{align}
\langle \calF | \hat{\scrS}_{S}\hat{W}_M | \calG \rangle
=\sum_{R_{A},R_{B}} N_{R_{B}S}^{\quad R_{A}} \bar{\calE}^{(S)}_{R_A} \calF_{R_A}(\{b\})
\calG_{R_B}(\{c\}),
\label{eq:coexist_surface_line_2_rep_basis}
\end{align}
respectively.
Note that 4D Wilson loops act as ``difference operators" and 4D surface defects do as diagonal multiplications in this basis.

When $M=\wedge^\ell \Box$ and $S=\wedge^k \Box$,
we can also repeat the similar computation to the previous one.
First of all, let us rewrite the eigenvalue $\bar{\calE}^{(S)}_{R}$ by using ~\eqref{eq:eigenvalues of difference operators} into
\begin{align}
\bar{\calE}^{(S)}_{R}=\sum_{L} \frq^{-2 \sum_{\underline{j} \in L}\subul{(\rho+\lambda_R)}{j}}
=\sum_{L} \frq^{-2 (\rho+\lambda_R,h_L)}
\end{align}
where $L$ runs over all the $\ell$-element subsets of $\{1,2,\ldots,N\}$.
Next, the sum including the Littlewood-Richardson coefficient can be written as follows.
\begin{align}
\sum_{R_{A},R_{B}} N_{R_{B}\wedge^k \Box}^{\qquad R_{A}}
=\sum_{\lambda_{R_B}} \sum_{K}
\end{align}
where $\displaystyle \lambda_{R_A}-\lambda_{R_B}=\sum_{i \in K} h_i=:h_K \in \Pi(\wedge^k \Box)$ and $K$ runs over all the $k$-element subsets of $\{1,2,\ldots,N\}$.

Now ~\eqref{eq:coexist_surface_line_2_rep_basis} leads to
\begin{align}
\langle \calF | \hat{\scrS}_{S}\hat{W}_M | \calG \rangle
&=\sum_{\lambda_B} \sum_{K,L}
\frq^{-2 (\rho+\lambda_B+h_K,h_L)}
\calF_{R(\lambda_B+h_K)}(\{b\})\calG_{R(\lambda_B)}(\{c\}) \\
&=\sum_{s=0}^{\min(k,\ell)} \frq^{-2\left( s-\tfrac{k\ell}{N} \right)}
\sum_{\lambda_B \substack{>=0 \quad \\>=-h_K}}\sum_{\substack{K,L \\ |K \cap L| =s}} 
\frq^{-2 (\rho+\lambda_B,h_L)}
\calF_{R(\lambda_B+h_K)}(\{b\})\calG_{R(\lambda_B)}(\{c\}) \\
\end{align}
where we use $(h_K,h_L)=\displaystyle \sum_{(i,j)\in K \times L} (h_i,h_j)=|K \cap L|-\dfrac{k\ell}{N}$ in the second line and $\lambda \ge 0$ means that it is a dominant weight, that is to say, $\lambda_\alpha \ge 0$ for all $\alpha$.
By evaluating ~\eqref{eq:coexist_surface_line_1_rep_basis} in the same way, we have the similar correspondence
\begin{align}
\raisebox{-0.45\height}{\input{figure/skein_relation/OPE_network_puncture_component_rep.tikz}}
\longleftrightarrow
\sum_{\substack{K,L \\ |K \cap L| =s}} 
\delta_{\lambda_A-\lambda_B,h_K}
\frq^{-(2\rho+\lambda_{B}+\lambda_{A},h_L)}
\end{align}
which is the dual expression of the operator $\hat{\calO}^{(k,\ell)}_{s}$.

Setting $s=k=\ell$, we finally get the following one needed later soon.
\begin{align}
\raisebox{-0.45\height}{\input{figure/open/punctured_loop_rep.tikz}} \longleftrightarrow \sum_{K}
\delta_{\lambda_A-\lambda_B,h_K}
\frq^{-(2\rho+\lambda_{A}+\lambda_{B},\lambda_A-\lambda_B)}.
\label{eq:punctured edge in rep. basis}
\end{align}

\subsection{Modified formula}
\label{subsec:modified formula}

\begin{figure}[h]
\centering
\input{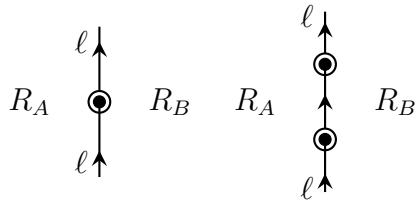}
\caption{Punctured edge.}
\label{fig:punctured edge}
\end{figure}

After performing the crossing resolutions, there appear several networks allowing the punctured edges as shown in Fig.~\ref{fig:punctured edge}.

The modification of the statistical model previously introduced in Sec.~\ref{subsec:proposal} is simple :
add another local Boltzmann factor for pairs of two adjacent dominant weights or, equivalently, edges.
The last result ~\eqref{eq:punctured edge in rep. basis} in the previous section suggests that this factor is given by
\begin{align}
B^{-,n}_{\lambda_A,\lambda_B}
=\frq^{-n(2\rho+\lambda_A+\lambda_B,\lambda_A-\lambda_B)}
=\frq^{-n(\wtl{\lambda}_A+\wtl{\lambda}_B,\wtl{\lambda}_A-\wtl{\lambda_B})}
\end{align}
where $n$ is the number of ``punctured" on the edge and $\wtl{\lambda}:=\lambda+\rho$.

As a simple application, we can see a new but naturally expected skein relation like
\begin{eqnarray}
\raisebox{-0.5\height}{\input{figure/open/junction_puncture_skein_1.tikz}}
=
\raisebox{-0.5\height}{\input{figure/open/junction_puncture_skein_2.tikz}}
\label{eq:junction_puncture_skein}
\end{eqnarray}
because of the following equality
\begin{align}
\frq^{-(\wtl{\lambda}_A+\wtl{\lambda}_C,\wtl{\lambda}_A-\wtl{\lambda}_C)}
=
\frq^{-(\wtl{\lambda}_A+\wtl{\lambda}_B,\wtl{\lambda}_A-\wtl{\lambda}_B)}
\frq^{-(\wtl{\lambda}_B+\wtl{\lambda}_C,\wtl{\lambda}_B-\wtl{\lambda}_C)}.
\end{align}

It is the interesting problem to prove the equalities ~\eqref{eq:OPE of network and puncture} or ~\eqref{eq:OPE of network and puncture2} based on the dual statistical model but we have no proof for them in general cases yet.

%% file: figure/open/punctured_loop.tikz
\tikzsetnextfilename{-figure-open-punctured_loop}
\begin{tikzpicture}
\def\l{1.0}
\def\r{0.08}
\coordinate (P0) at (0,0);
\coordinate (U0) at ($(P0)+(0,\l)$);
\coordinate (D0) at ($(P0)+(0,-\l)$);

\draw[mid-end arrow] (P0) -- node[above left]{$k$} (U0);
\draw[fill=white,thick] (P0) circle (1.8*\r);
\draw[fill=black] (P0) circle (\r);
\draw[start-mid arrow] (D0) -- node[below left]{$k$} (P0);
\end{tikzpicture}

%% file: figure/open/punctured_loop_rep.tikz
\tikzsetnextfilename{-figure-open-punctured_loop_rep}
\begin{tikzpicture}
\def\l{1.0}
\def\r{0.08}
\coordinate (P0) at (0,0);
\coordinate (U0) at ($(P0)+(0,\l)$);
\coordinate (D0) at ($(P0)+(0,-\l)$);

\draw[mid-end arrow] (P0) -- node[above left]{$k$} (U0);
\draw[fill=white,thick] (P0) circle (1.8*\r);
\draw[fill=black] (P0) circle (\r);
\draw[start-mid arrow] (D0) -- node[below left]{$k$} (P0);
\draw[thick] ($(P0)+(-0.5,0)$) node[left]{$R_A$};
\draw[thick] ($(P0)+(0.5,0)$) node[right]{$R_B$};
\end{tikzpicture}

%% file: figure/open/junction_puncture_skein_1.tikz
\tikzsetnextfilename{-figure-open-junction_puncture_skein_1}
\begin{tikzpicture}
\def\l{1.0}
\def\r{0.08}
\coordinate (O) at (0,0);
\coordinate (PA) at ($(O)+(120:\l)$);
\coordinate (A) at ($(O)+(120:2*\l)$);
\coordinate (B) at ($(O)+(0:2*\l)$);
\coordinate (C) at ($(O)+(240:2*\l)$);
\coordinate (AB) at ($(A)!0.5!(B)$);
\coordinate (BC) at ($(B)!0.5!(C)$);
\coordinate (CA) at ($(C)!0.5!(A)$);

\draw[mid-end arrow] (PA) -- (O);

\draw[fill=white,thick] (PA) circle (1.8*\r);
\draw[fill=black] (PA) circle (\r);

\draw[mid arrow] (B) -- node[below right]{$b$} (O);
\draw[mid arrow] (C) -- node[left]{$c$} (O);
\draw[start-mid arrow] (A) -- node[left]{$a$} (PA);

\draw[thick] (AB) node[above right]{$R_A$};
\draw[thick] (BC) node[below right]{$R_B$};
\draw[thick] (CA) node[left]{$R_C$};
\end{tikzpicture}

%% file: figure/open/junction_puncture_skein_2.tikz
\tikzsetnextfilename{-figure-open-junction_puncture_skein_2}
\begin{tikzpicture}
\def\l{1.0}
\def\r{0.08}
\coordinate (O) at (0,0);
\coordinate (A) at ($(O)+(120:2*\l)$);
\coordinate (PB) at ($(O)+(0:\l)$);
\coordinate (B) at ($(O)+(0:2*\l)$);
\coordinate (PC) at ($(O)+(240:\l)$);
\coordinate (C) at ($(O)+(240:2*\l)$);
\coordinate (AB) at ($(A)!0.5!(B)$);
\coordinate (BC) at ($(B)!0.5!(C)$);
\coordinate (CA) at ($(C)!0.5!(A)$);

\draw[start-mid arrow] (B) -- node[below right]{$b$} (PB);
\draw[start-mid arrow] (C) -- node[left]{$c$} (PC);
\draw[mid arrow] (A) -- node[left]{$a$} (O);

\draw[fill=white,thick] (PB) circle (1.8*\r);
\draw[fill=black] (PB) circle (\r);
\draw[fill=white,thick] (PC) circle (1.8*\r);
\draw[fill=black] (PC) circle (\r);

\draw[mid-end arrow] (PB) -- (O);
\draw[mid-end arrow] (PC) -- (O);

\draw[thick] (AB) node[above right]{$R_A$};
\draw[thick] (BC) node[below right]{$R_B$};
\draw[thick] (CA) node[left]{$R_C$};
\end{tikzpicture}

%% file: appendix.tex
\section{More mathematics on the dual model}
\label{app:more math}
%alternative equivalent algorithm}

Here we develop some useful tools to compute the local Boltzmann factor $B^\triangle$ using ~\eqref{eq:final exp. for local Boltzmann factor} and to prove some skein relations in App.\ref{app:derivation of skeins}.
First of all, recall the notations of Lie algebras and their representations.
Consider the case that the Lie algebra is $\Liesu (N)$.
$R(\lambda)$ denotes the irreducible representation associated with a dominant weight $\lambda$, $\lambda_R$ does the dominant weight to $R$ conversely and $\Pi(R)$ does the set of weights in $R$.
$\omega_\alpha$ for $\alpha =1,2,...,N-1$ are fundamental weights, $h_i$ for $i=1,2,\ldots,N$ are weights in $\Pi(R(\omega_1)=\Box)$
\footnote{The perfect order of the indices of $h_i$ is determined by the partial order in the weight lattice.}
and there are relations between two as $h_a=\omega_a-\omega_{a-1}$ where $\omega_N=\omega_0=0$.
We also use the standard metric in weight vectors determined by $\displaystyle h_i=e_i - \tfrac{1}{N} \sum_{i=1}^N e_i$ and $(e_i,e_j)=\delta_{i,j}$.

\subsection{Definitions}

Let us start by repeating some definitions which appeared in Sec.~\ref{subsubsec:formula for local Boltzmann factor}.

We introduced a mathematical object which we call ``pyramid".
This is just an assembly of integers designated by two labels $h$ and $\alpha=\alpha_h$.
$h$ runs over $1,2,\ldots,N-1$ and $\alpha$ does over $1,2,\ldots,N-h$ for each $h$. Therefore, this object consists of $\tfrac{1}{2}N(N-1)$ integers.
There is an inclusion of weights into the pyramid as follows :
\begin{align}
\hat{\lambda}_{h:\alpha} :=\sum_{\beta=\alpha}^{\alpha+h-1} \lambda_\beta
\end{align}
where $\displaystyle \lambda=\sum_{\beta=1}^{N-1} \lambda_\beta \omega_\beta$.
We also use the same symbol $\hat{\lambda}$ for the pyramids not in the image of this inclusion map. In such cases, $\hat{\lambda}$ is considered as a single symbol as a whole and $\lambda$ is meaningless.
Note that the addition can be defined as
\begin{align}
(c_1 \hat{s}_1 + c_2 \hat{s}_2)_{h:\alpha}:=c_1(\hat{s}_1)_{h:\alpha}+c_2(\hat{s}_2)_{h:\alpha}
\end{align}
which is consistent with the above inclusion map in the sense that it preserves the original additional structure in the weight vector space.
$\hat{0}$ is the identity element of this operation.
There can be also a product defined as
\begin{align}
(\hat{s}_1 * \hat{s}_2)_{h:\alpha}:=(\hat{s}_1)_{h:\alpha}(\hat{s}_2)_{h:\alpha}.
\end{align}
The distributive property is obvious.

We also defined two functions :
\begin{enumerate}
\item ~\eqref{eq:majority} majority function $mj$ for three variables :
\begin{align}
mj(a,b,c):=
\begin{cases}
a \qquad b=a \;\textrm{or}\; c=a \\
b \qquad a=b \;\textrm{or}\; c=b \\
c \qquad a=c \;\textrm{or}\; b=c
\end{cases}
\end{align}
and
\begin{align}
mj(\hat{\lambda_1},\hat{\lambda_2},\hat{\lambda_3}):=\{mj((\hat{\lambda_1})_{h:\alpha},(\hat{\lambda_2})_{h:\alpha},(\hat{\lambda_3})_{h:\alpha})\}_{h:\alpha}
\end{align}
As there appears no case that all variables are distinct, this definition is well-defined in our usage

\item ~\eqref{eq:dimension of pyramid} $q$-dimension function $D$ :
\begin{align}
D[\hat{\lambda}]:=\prod_{h=1}^{N-1}\prod_{\alpha=1}^{N-h}\dfrac{[(\hat{\lambda})_{h:\alpha}+h]_q}{[h]_q}
\end{align}
\end{enumerate}
and there is a simple relation to the ordinary $q$-dimension as
\begin{align}
\dim_q R(\lambda)=D[\hat{\lambda}]
\label{eq:q-dimension}
\end{align}
where $\hat{\lambda}$ is the natural inclusion into the pyramid of the dominant weight $\lambda$.

Finally, let us introduce following pyramids defined for any two subsets $I,J$ of $\{1,2,\ldots,N\}$ satisfying $I \cap J=\phi$ :
\begin{align}
\hat{f}_{I,J}:=mj(\hat{0},-\hat{h_I},\hat{h_J})
\end{align}
where $\displaystyle h_I:=\sum_{i \in I} h_i$.
They have equivalent definitions
\begin{align}
(\hat{f}_{I,J})_{h:\alpha}:=
\begin{cases}
+1 \quad & {\rm if}\;\; \alpha \in J \;{\rm and}\; \alpha+h \in I \\
-1 \quad & {\rm if}\;\; \alpha \in I \;{\rm and}\; \alpha+h \in J \\
0 \quad & {\rm otherwise}
\end{cases}
\end{align}
or
\begin{align}
(\hat{f}_{I,J})_{h:\alpha}:=
\sum_{i \in I,j\in J}
{\rm sgn}(i-j)
\delta_{h,|i-j|}\delta_{h,\min(i,j)}
\end{align}
where $\delta$ is the ordinary Kronecker's $\delta$ symbol.

These pyramids satisfy the following properties :
\begin{align}
\hat{f}_{I,J}&=-\hat{f}_{J,I} \qquad {\rm skewsymmetric} \\
\hat{f}_{I,J\sqcup K}&=\hat{f}_{I,J}+\hat{f}_{I,K} \qquad {\rm linearity} \\
\hat{h_J}&=\hat{f}_{\bar{J},J}.
\end{align}

\subsection{Convenient formulae}

Now let us start the argument recalling the discussion in Sec.\ref{subsubsec:allowed}.
Consider three regions called $A$,$B$ and $C$ clockwise around a trivalent junction and denote their dominant weights $\lambda_A$, $\lambda_B$ and $\lambda_C$. See Fig.\ref{fig:region and junction} in Sec.\ref{subsubsec:allowed}. We also denote the outgoing charge associated with the edge between the regions $X$ and $Y$ by $a_{XY}$ for $(X,Y)=(A,B),(B,C)$ and $(C,A)$.

We define the following objects in order.
\begin{align}
\lambda_{XY}:=\lambda_X-\lambda_Y=:\sum_{\alpha=1}^{N-1} \Lambda^{\alpha}_{XY}\omega_\alpha=:\sum_{s=1}^{N} \lambda^{s}_{XY}h_s.
\label{eq:lambda_XY expansion}
\end{align}
$\lambda^{s}_{XY}$ is not uniquely determined due to the condition $\displaystyle \sum_s h_s=0$ in the root vector space.
But it is uniquely determined if we impose the conditions
$\lambda^{s}_{XY} \ge 0$ and $\exists s\; \lambda^{s}_{XY}=0$.

We can find that $\lambda^{s}_{XY}$ is either $1$ or $0$ and define $E_{XY}:=\{ s \in \{1,2,\ldots,N\} | \lambda^{s}_{XY}=1 \}$ where $|E_{XY}|=a_{XY}$ follows and $\overline{E_{XY}}:=\{1,2,\ldots,N\} \backslash E_{XY}=E_{YX}$.
The cycle condition $\lambda_{AB}+\lambda_{BC}+\lambda_{CA}=0$ tells us $E_{AB} \sqcup E_{BC} \sqcup E_{CA}=\{1,2,\ldots,N\}$
(disjoint union).
Now we have
\begin{align}
mj(\lambda_A,\lambda_B,\lambda_C)=\hat{\lambda_A}+\hat{f}_{E_{AB},E_{CA}}
\label{eq:another expression of majority function}
\end{align}
and we obtain two other similar expressions by permuting the above cyclically as $A \to B \to C \to A$.
This formula will turn out to be useful in the next section.

Finally, we list a few propositions also used later.
\begin{enumerate}
\item
\begin{align}
D[\hat{x}+\hat{z}]D[\hat{y}]=D[\hat{x}]D[\hat{y}+\hat{z}]
\;\;\textrm{when}\;\; (\hat{x}-\hat{y})*\hat{z}=\hat{0}.
\label{prop:DD=DD}
\end{align}
Each element in pyramid satisfies $x=(\hat{x})_{h:\alpha}=(\hat{y})_{h:\alpha}=y$ or $z=(\hat{z})_{h:\alpha}=0$ and then we can say $[x+z]_q[y]_q=[x]_q[y+z]_q$ for any $(h:\alpha)$.\\

\item 
\begin{align}
\hat{f}_{I,J} *\hat{f}_{K,L}=\hat{0} \for (I \sqcup J) \cap (K \sqcup L)=\phi
\label{prop:2nd}
\end{align}
Using $(\hat{f}_{P,Q})_{h:\alpha}=0$ for $\alpha \notin P \sqcup Q$, this statement holds true.\\

\item 
\begin{align}
\hat{f}_{I,J} * \hat{f}_{I,K}=\hat{0} \for J \cap K = \phi
\label{prop:3rd}
\end{align}

Assume $(\hat{f}_{I,J})_{h:\alpha}\neq 0$ and $(\hat{f}_{I,K})_{h:\alpha}\neq 0$ for some common $(h:\alpha)$.
If $\alpha \in I$, we have $\alpha+h \in J$ and $\alpha+h \in K$ but it is impossible by definition and we say $J \cap K=\phi$. This is same for the case $\alpha+h \in I$.
So the assumption is always false and the above statement is true.
\end{enumerate}

Note that there is a more general formula including last two propositions :
\begin{align}
\hat{f}_{I,J}*\hat{f}_{K,L}=(\hat{f}_{I \cap K,J \cap L})^2 -(\hat{f}_{I \cap L,J \cap K})^2
\end{align}
where $\hat{x}^2:=\hat{x}*\hat{x}$.

\section{Derivation of several skein relations}
\label{app:derivation of skeins}

Based on our proposal for the local Boltzmann factor ~\eqref{eq:final exp. for local Boltzmann factor},
we prove elementary skein relations in this appendix.

\subsection{Associativity}

The associativity skein relation is given by
\begin{eqnarray}
\raisebox{-0.5\height}{\input{figure/app/s-channel.tikz}}
=
\raisebox{-0.5\height}{\input{figure/app/t-channel.tikz}}\label{eq:flip}.
\end{eqnarray}
Here $s=p+q=v-r$, $t=q+r=v-p$ and $v=p+q+r$.
Their dual triangle quivers are given as
\begin{eqnarray}
\raisebox{-0.5\height}{\input{figure/app/s-channel_triangles.tikz}}
=
\raisebox{-0.5\height}{\input{figure/app/t-channel_triangles.tikz}}\label{eq:flip_triangle}.
\end{eqnarray}
which is the flip of the triangulations. We have introduced here $P:=E_{CB}$, $Q:=E_{BA}$ and $R:=E_{AD}$.
~\eqref{eq:another expression of majority function} tells that the local Boltzmann factors' expression associated with this equality is equivalent to
\begin{align}
D[\hat{\lambda_B}+\hat{f}_{Q,P}]^{-1/2}D[\hat{\lambda_A}+\hat{f}_{R, P\sqcup Q}]^{-1/2}=D[\hat{\lambda_B}+\hat{f}_{Q\sqcup R,P}]^{-1/2}D[\hat{\lambda_A}+\hat{f}_{R,Q}]^{-1/2}
\end{align}
for any $\lambda_A,\lambda_B,\lambda_C$ and $\lambda_D$. In the following, we prove this equality.

Introduce $\hat{x}:=\hat{\lambda_B}+\hat{f}_{Q,P}$ and $\hat{y}:=\hat{\lambda_A}+\hat{f}_{R,Q}$. Now we get
\begin{align}
(l.h.s)^{-2}=D[\hat{x}]D[\hat{y}+\hat{f}_{R,P}] \\
(r.h.s)^{-2}=D[\hat{x}+\hat{f}_{R,P}]D[\hat{y}].
\end{align}
To apply the proposition ~\eqref{prop:DD=DD} to the above,
it is enough to check $(\hat{x}-\hat{y})*\hat{f}_{R,P}=\hat{0}$.
Since $\hat{\lambda_B}-\hat{\lambda_A}=\hat{h_Q}=\hat{f}_{\bar{Q},Q}$,
\begin{align}
& (\hat{x}-\hat{y})*\hat{f}_{R,P}=
(\hat{f}_{\bar{Q},Q}+\hat{f}_{Q,P}+\hat{f}_{R,Q})*\hat{f}_{R,P} \\
&=(2\hat{f}_{R,Q}+\hat{f}_{\overline{P \sqcup Q \sqcup R},Q})*\hat{f}_{R,P} = \hat{0}
\end{align}
where we have used the two propositions ~\eqref{prop:2nd} and ~\eqref{prop:3rd} in the last line. Now we have proved the expected equality.
%Note that the skein relation of triangle contraction shown in Sec.3 in \cite{TW1504} automatically from the associativity.

\subsection{Digon contractions}

There is more non-trivial skein relations what we call digon contractions as shown below.
\begin{eqnarray}
\raisebox{-0.45\height}{\input{figure/app/digon.tikz}}
=
\dfrac{[a+b]_q!}{[a]_q![b]_q!}
\raisebox{-0.45\height}{\input{figure/app/digon_contract.tikz}}
\label{eq:digon_contract}
\end{eqnarray}
where $[n]_q!:=\prod_{i=1}^{n} [i]_q$ for a positive integer $n$.

First of all, let us introduce several definitions.
For fixed $E_{AB}$, we define a natural embedding
\begin{align}
& \ell^{-1}: \{1,2,\ldots,a+b\} \overset{bijec.}{\longrightarrow} E_{AB} \\
& \ell^{-1}_\gamma:=\ell^{-1}(\gamma) \for \gamma \in \{1,2,\ldots a+b\}
\qquad \ell_i:=\ell(i) \for i \in E_{AB}
\end{align}
satisfying that $1\le \ell^{-1}_{\gamma} <\ell^{-1}_{\gamma'}\le N$ for $1\le \gamma<\gamma' \le a+b$.
\begin{align}
M:=\{ (\ell^{-1}_{\gamma'}-\ell^{-1}_\gamma,\ell^{-1}_\gamma) \for {\rm any}\; \gamma'>\gamma \}.
\end{align} 
By definition, for any two subsets $I$,$J$ ($I \cap J=\phi$) of $E_{AB}$, it is true that $(\hat{f}_{I,J})_{h:\alpha} \neq 0 \Longleftrightarrow (h,\alpha) \in M$.

Next, let $\check{M}$ be the index set of the pyramid for $SU(a+b)$ weights. In other words, for $(\check{h}:\check{\alpha}) \in \check{M}$, $\check{h}$ runs over $1$ to $a+b-1$ and $\check{\alpha}$ does over $1$ to $a+b-\check{h}$.
The map $\ell$ induces a new bijection map $\check{\ell}$ from $M$ to $\check{M}$ as follows.
\begin{align}
(\check{h}:\check{\alpha}):=\check{\ell}(h,\alpha):=
(\ell_{\alpha+h}-\ell_{\alpha}:\ell_{\alpha})
\end{align}

If we label the representation assigned to the inside region of the digon as $S$, the left hand side gives
\begin{align}
&\sum_{S} \dfrac{\dim_q S}{D[mj(\hat{\lambda_A},\hat{\lambda_B},\hat{\lambda_S})]}
=
\sum_{I:=E_{SB} \subset E_{AB}} \dfrac{D[\hat{\lambda_B}+\hat{h_I}]}{D[\hat{\lambda_B}+\hat{f}_{E_{BA},I}]} \\
=&\sum_{\substack{I \subset E_{AB}\\ |I|=b}} \dfrac{D[\hat{\lambda_B}+\hat{f}_{E_{BA},I}+\hat{f}_{E_{AS},I}]}{D[\hat{\lambda_B}+\hat{f}_{E_{BA},I}]}
=\sum_{\substack{I \subset E_{AB}\\ |I|=b}} \dfrac{D[\hat{\lambda_B}+\hat{f}_{E_{AS},I}]}{D[\hat{\lambda_B}]} \\
=&\sum_{\substack{I \subset E_{AB}\\ |I|=b}} \prod_{(h,\alpha)\in M} \dfrac{[(\hat{\lambda_B}+\hat{f}_{E_{AS},I})_{h:\alpha}+h]_q}{[\hat{\lambda_B}+h]_q}
=\sum_{\substack{I \subset E_{AB}\\ |I|=b}} \prod_{(\check{h},\check{\alpha})\in \check{M}} \dfrac{[(\hat{\mu})_{\check{h}:\check{\alpha}}+(\hat{f}_{E_{AS},I})_{\check{\ell}^{-1}(\check{h},\check{\alpha})}+\check{h}]_q}{[(\hat{\mu})_{\check{h}:\check{\alpha}}+\check{h}]_q}
\label{eq:digon_lhs}
\end{align}
where we have used $h_I=\hat{f}_{\bar{I},I}$, $\bar{I}=E_{BA} \sqcup E_{AS}$ and the proposition ~\eqref{prop:DD=DD} using also \eqref{prop:3rd} $\hat{f}_{E_{BA},I}*\hat{f}_{E_{AS},I}=\hat{0}$.
In the 3rd line, we have used $(\hat{f}_{E_{AS},I})_{h:\alpha}=0$ for $(h,\alpha) \notin M$ and $M \overset{\check{\ell}}{\simeq} \check{M}$ and redefined $(\hat{\mu})_{\check{h}:\check{\alpha}}:=(\hat{\lambda_B})_{\check{\ell}^{-1}(\check{h},\check{\alpha})}+h-\check{h}$ where $h=h(\check{h},\check{\alpha})=\ell^{-1}_{\check{h}+\check{\alpha}}-\ell^{-1}_{\check{\alpha}}$.

Now what we should prove are two following equations.
\begin{align}
&\sum_{\beta=\check{\alpha}}^{\check{\alpha}+\check{h}-1} \hat{\mu}_{1:\beta}=\hat{\mu}_{\check{h}:\check{\alpha}} \\
&(\widehat{h^{SU(a+b)}_{\ell(I)}})_{\check{h},\check{\alpha}}=(\hat{f}_{E_{AS},I})_{\check{\ell}^{-1}(\check{h},\check{\alpha})}
\end{align}
The former equality says that $\hat{\mu}$ is the image of a weight $\mu$ in the pyramid and follows from the direct computation based on the above definitions.
The latter one means that $\hat{f}_{E_{AS},I}$ gives an image of a weight in $\Pi(\wedge^{b}\Box)$ of $SU(a+b)$, and it also readily follows from the equality
\begin{align}
(\widehat{h^{SU(a+b)}_{\ell(I)}})_{\check{h},\check{\alpha}}=(\hat{f}_{E_{AB}\backslash I,I})_{\ell^{-1}_{\alpha+h}-\ell^{-1}_\alpha:\ell^{-1}_\alpha}.
\end{align}

In conclusion, the numerator in ~\eqref{eq:digon_lhs}
equals to the $q$-dimension of the $SU(a+b)$ irreducible representation $R(\mu+h_{\ell(I)})$ up to the common factor $\prod_{(\check{h}:\check{\alpha})} [\check{h}]_q$
and, the sum over all the $b$ element subsets of $E_{AB}$ equals to all irreducible representations appearing in the tensor product of $R(\mu)$ and $\wedge^{b}\Box$.
Therefore, this gives $\dim_q^{SU(a+b)} \wedge^b\Box$ which exactly reproduces the prefactor in the right hand of ~\eqref{eq:digon_contract}.

\section{General charge/network correspondence}
\label{app:charge_network}

This appendix is a complement of the paper \cite{TW1504}.
There we see the one-to-one mapping between the charge lattice for $\SUSYN{4}$ $\Liesu(3)$ theory proposed by Kapustin \cite{Kapustin05} and $A_2$ networks on the 2-torus and also show several examples for $A_{N-1}$ cases in the appendix.
However, we did not explain the dictionary in detail.
Here we state the mapping for general $A_N$.
This is a minimal extension of a work for general $A_1$ class S theories \cite{DMO09}.
\footnote{Here we consider $\SUSYN{4}$ SYM as the very special case of class S theories. The similar relations are expected to hold true in the $\SUSYN{2}^\ast$ gauge theory but precise dictionaries are not established completely because there appears a flavor symmetry related to the hypermultiplet mass term.}
The generalization and refinement to general class S theories are interesting future problems.
%Hereafter, we can discuss in the classical limit $q \to 1$ or $\frq \to 1$ and take this limit.
Note that the following relations can hold true in the Liouville-Toda CFTs and also that the expectation values vanish in the 2D $q$-deformed Yang-Mills when their electric/magnetic weights are not in the root lattices as explained in part 3 in Sec.~\ref{subsec:review on qYM}.

\subsection{Useful symbol}

\begin{figure}[h]
\centering
\input{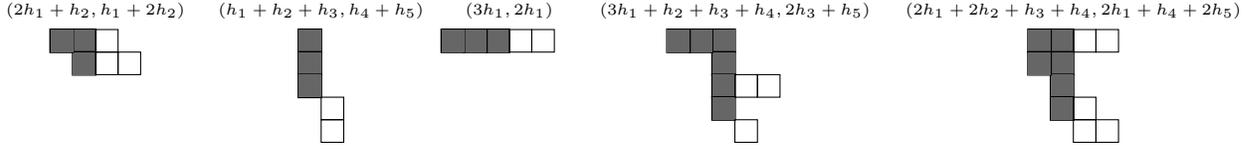}
\caption{Several examples for the relation between an element of the $\Liesu(N)$ charge lattice (above) and its diagrammatic symbol (below).}
\label{fig:symbols}
\end{figure}

Here we introduce a useful symbol expressing an element of $\Liesu(N)$ Wilson-'t Hooft loop charge lattice $( \Lambda_{mw} \times \Lambda_{wt} ) / \calW_{\Liesu(N)}$ where $\Lambda_{mw}$, $\Lambda_{wt}$ and $\calW_{\Liesu(N)}$ are the magnetic weight lattice, the weight lattice and the Weyl reflection group, respectively. Be aware that $\Lambda_{mw} \simeq \Lambda_{wt}$ for $\Liesu(N)$ and then we use the same basis.
For a given pair of $(\mu,\lambda)$, it is always possible to take $\mu$ into a dominant weight $\mu'$ using a Weyl reflection. According to this operation, $\lambda$ is also mapped into an element $\lambda'$ which is not always uniquely determined.
There, we have a Young diagram $Y_M$ associated with $\mu'$.
In the same way as \eqref{eq:lambda_XY expansion},
$\lambda'$ can be expanded with $h_{s}$ and we have unique elements $\lambda'^{\subul{s}}$ ($s=1,2,\ldots,N$) which are non-negative integers.
By putting $\lambda'^{\subul{s}}$ boxes in the $s$-th row in the similar way as the ordinary Young diagrams, we have a diagram referred to as $Y_E$.
Now, we make a new diagram which is a pair of the horizontally flipped and filled $Y_M$ and the diagram $Y_E$.
See Fig.~\ref{fig:symbols} below for examples.

\subsection{Charge to network}

For a given charge pair $(\mu,\lambda)$,
let $M_i$ be subsets of ${1,2,\ldots,N}$ so that there is a box in $Y_M$ specified by $i$-th column and $a$-th row only if $a \in M_i$.
By replacing $Y_M$ by $Y_E$, we also define $E_i$ in the same way.
Then, define $s^{pq}$ as the number of elements of $M_{p} \cap E_{q}$ and $Q_{(s)}^{pq}$ as an open network like
\begin{align}
Q_{(s)}^{pq} \longleftrightarrow \raisebox{-0.5\height}{\input{figure/app/Q_on_torus.tikz}}.
\end{align}
By using these, the BPS Wilson-'t Hooft line operator in $\SUSYN{4}$ $SU(N)$ SYM is geometrically represented by
\begin{align}
\input{figure/app/full_network.tikz}
\end{align}
where edges are connected on any adjacent parallelograms and each pair of opposite edges is identified.
Note also that this relation holds up to lower charges (see the beginning of Sec.4 in \cite{TW1504}).
We show two examples.
\begin{align}
\raisebox{-0.5\height}{\input{figure/app/A_diag.tikz}}
\longleftrightarrow\raisebox{-0.5\height}{\input{figure/app/A.tikz}}
\qquad
\raisebox{-0.5\height}{\input{figure/app/B_diag.tikz}}
\longleftrightarrow\raisebox{-0.5\height}{\input{figure/app/B.tikz}}.
\qquad
\end{align}

The reversed operation can be done by computing the trace functions associated with the network because the trace function is a polynomial (allowing negative powers) of two $U(1)^{N}/U(1)$ fugacities along $\alpha$-cycle and $\beta$-cycle of the 2-torus.

\begin{comment}
\subsection{Network $\to$ charge}

The charges associated with a network can be read off from its trace function.
Although the similar discussion has already appeared in \cite{CGT15}, we show another treatment based on \cite{CKM12} (See also Sec.3.3 in \cite{TW1504} for our usage), for example, in the specific case.
Let cut the 2-torus along a given pair of $\alpha$-cycle and $\beta$-cycle.
Then, any network on it can viewed as an endomorphism of some vector spaces.
Such vector spaces are given as follows.
With each intersecting point of an edge with charge $k$ with $\alpha$ cycle or $\beta$ cycle, we assign a $\wedge^{k} \Box$ representation space $V$ or $W$, respectively.
The tensor product of these spaces is the answer.
Since the order of the tensor product is irrelevant in the classical limit, we do not pay attentions to that.
Now, we can define a 
\end{comment}

%% file: figure/app/digon_contract.tikz
\tikzsetnextfilename{-figure-app-digon_contract}
\begin{tikzpicture}
\def\w{2.8}
\def\h{2.4}
\coordinate (S) at (0,0);
\coordinate (T) at ($(S)+(\h,0)$);
\coordinate (MU) at (0.5*\h,0.15*\w);
\coordinate (MD) at (0.5*\h,-0.15*\w);

\draw[mid arrow] (S) -- node[above]{$a+b$} (T);

\draw[thick] (MU) node[above]{$R_A$};
\draw[thick] (MD) node[below]{$R_B$};
\end{tikzpicture}